\documentclass[lettersize,journal]{IEEEtran}
\usepackage{amsmath,amssymb,amsthm,graphicx,subcaption}
\usepackage{mathrsfs,algorithm,algpseudocode}
\usepackage{booktabs,multirow,bm,diagbox}
\usepackage{cite}
\usepackage{xcolor}
\usepackage{mathtools}
\usepackage{framed}
\graphicspath{{figures/}}

\begin{document}

\onecolumn 
\pagestyle{empty} 
\begin{framed}
This work has been submitted to the IEEE for possible publication. Copyright may be transferred without notice, after which this version may no longer be accessible. 2023 IEEE. Personal use of this material is permitted. Permission from IEEE must be obtained for all other uses, in any current or future media, including reprinting/republishing this material for advertising or promotional purposes, creating new collective works, for resale or redistribution to servers or lists, or reuse of any copyrighted component of this work in other works.
\end{framed}
\twocolumn 
\setcounter{page}{1} 
\pagestyle{headings} 

\newtheorem{theorem}{Theorem}
\newtheorem{exmp}{Example}
\newtheorem{lemma}{Lemma}
\newtheorem{proposition}{Proposition}
\newtheorem{definition}[exmp]{Definition}

\renewcommand{\algorithmicrequire}{\textbf{Input:}}
\renewcommand{\algorithmicensure}{\textbf{Output:}}
\newcommand{\tabincell}[2]{\begin{tabular}{@{}#1@{}}#2\end{tabular}}

\title{\emph{Probe}: Learning Users' Personal Projection Bias in Inter-temporal Choices}

\author{Qingming Li, \quad  H. Vicky Zhao
\thanks{This work was supported by Youth Research Project of Zhejiang Lab (No. K2023KG0AA04). \emph{(Corresponding author: H. Vicky Zhao)}}%
\thanks{Qingming Li is with the Research Institute of Artificial Intelligence, Zhejiang Lab, Hangzhou 311100, China (email: liqm@zhejianglab.com).}%
\thanks{H. Vicky Zhao is with the Department of Automation, Beijing National Research Center for Information Science and Technology, Tsinghua University, Beijing 100084 P. R. China (email: vzhao@tsinghua.edu.cn).}}%

\maketitle

\begin{abstract}
Inter-temporal choices involve making decisions that require weighing costs in the present against benefits in the future. One specific type of inter-temporal choice is the decision between purchasing an individual item at full price or opting for a bundle including that item at a discounted price. Previous works assume that individuals have accurate expectations of factors involved in these decisions. However, in reality, users' perceptions of these factors are often biased, leading to irrational and suboptimal decision-making. In this work, we specifically focus on two commonly observed biases: the projection bias and the reference-point effect. To address these biases, we propose a novel bias-embedded preference model called \emph{Probe}. \emph{Probe} introduces prospect theory from behavioral economics to model these biases, which incorporates a weight function to capture users' projection bias and a value function to account for the reference-point effect. This allows us to determine the probability of users selecting the bundle or a single item. We theoretically analyze the impact of projection bias on bundle pricing strategies. Through experimental results, we show that the proposed \emph{Probe} model outperforms existing methods and leads to a better understanding of users' behaviors in bundle purchases. This investigation can facilitate a deeper comprehension of users' decision-making processes, enable personalized services, and help sellers design better product bundling strategies.
\end{abstract}








\section{Introduction}
\label{sec:intro}
\emph{Inter-temporal choices} happen frequently in everyday life, which refer to decisions involving tradeoffs between cost in the present and benefits in the future. For example, how much food to buy for tomorrow's lunch, or whether to take out a health insurance policy to avoid future potential loss. To make inter-temporal choices, users typically consider a variety of factors, including the probability of future outcomes and the cost of each outcome. However, a large stream of literature suggests that users' perceptions of these factors are usually biased \cite{aggarwal2019modeling}. These biases may lead users to irrational and suboptimal decisions, such as buying too much food when going shopping with an empty stomach, or more likely to buy health insurance when the air quality is worse 
\cite{chang2018something}.  
It is essential to identify these biases and better model users' inter-temporal choices. This study can help understand users' decision-making mechanisms, enable personalized services, and design effective pricing and marketing strategies.

\begin{figure*}[tbp]
\centering
\begin{minipage}[b]{.23\linewidth}
  \centering
  \centerline{\includegraphics[width=3.8cm]{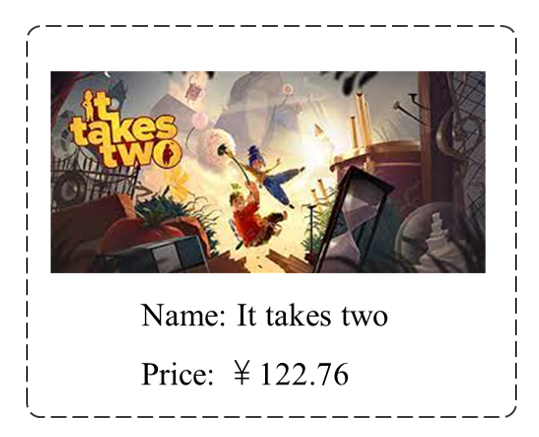}}
  \centerline{(a) the single game}\medskip
\end{minipage}
\hfill
\begin{minipage}[b]{.75\linewidth}
  \centering
  \centerline{\includegraphics[width=11cm]{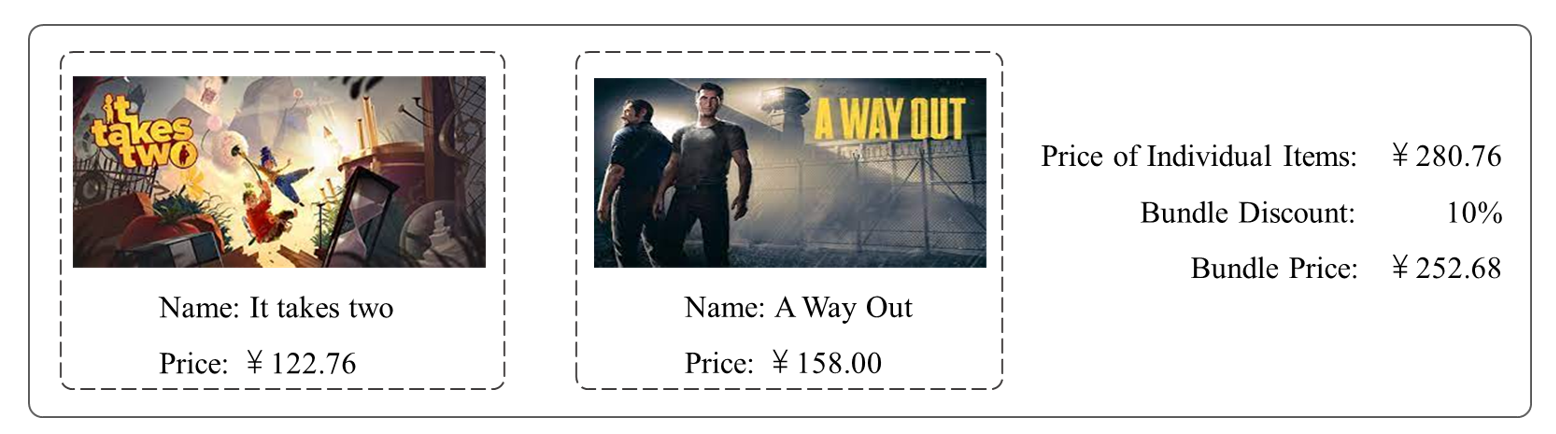}}
  \centerline{(b) the bundle}\medskip
\end{minipage}
\caption{The inter-temporal choice in the bundle sale scenario.}
\label{fig:bundle_choice}
\end{figure*}

In our study, we investigate inter-temporal choice modeling, with a specific focus on users' purchasing behaviors in the context of bundle sales. We call the problem studied in this work the \emph{bundle choice problem}. Consider a user who intends to purchase an individual game, as shown in Fig.~\ref{fig:bundle_choice}(a). Meanwhile, the platform reminds the user that a bundle containing this game is available for sale with a discount, as shown in Fig.~\ref{fig:bundle_choice}(b). Typically, the user has to decide whether to buy the individual game at full price or purchase the bundle at a reduced price. Note that the user has no current demand for the additional item, and it is not clear whether he/she will need the additional item in the future. In this example, if the user chooses to purchase the bundle, it may result in a waste of money if the user does not need the additional item in the future; while if the user chooses to purchase the individual item, no discount will be given to the additional item if the user does need it in the future. Consequently, in the bundle choice problem, the user needs to consider the trade-off between two factors: the additional cost to obtain the bundle at present and the probability of needing the additional item in the future.

Prior works in \cite{loewenstein2003projection,tversky1993context} have demonstrated the presence of biases in users' perceptions regarding the two aforementioned factors. The first type of bias is known as \emph{projection bias}, which manifests in users' estimates of the probability of needing the additional item in the future \cite{loewenstein2003projection}. Typically, users' preferences for the additional item undergo dynamic changes over time \cite{pujahari2022handling}, leading to a potential discrepancy between their current preference and their future preference. However, users with projection bias tend to underestimate this preference change and assume that their future preference will resemble their current preference. Consequently, their estimates of the above probability are not aligned with their actual future preferences, but rather based on their current preference. 
Projection bias can lead to underestimation or overestimation of the probability depending on the relative magnitudes of the current and future preferences \cite{kaufmann2022projection}. Specifically, when the current preference is lower than the actual future preference, users' estimated future preference and the estimates of the probability are correspondingly lower than the actual values, resulting in underestimation. Conversely, when the current preference is higher than the actual future preference, users' estimations are higher than actual values, leading to overestimation.

The second type of bias is the \emph{reference-point effect}, which arises from users' perceptions of price \cite{tversky1993context}. Specifically, users typically have an expected price for an item based on prior purchase experience or market research. The expected price serves as a reference point in their decisions. The satisfaction that users obtain from an item does not depend on the absolute value of its price, but rather on the difference between the actual price and users' expected price. If the actual price is lower than users’ expectations, then they perceive gains from the purchase. Otherwise, users perceive loss. 

\textbf{Limitations of Prior Works.} 
There have been many works in the literature trying to explore users' behaviors in bundle scenarios, while there may be subtle distinctions from the bundle choice problem studied in our work. Specifically, existing works related to the problem can be categorized into two main topics: bundle strategies and inter-temporal choice modeling. In bundle strategies, a common assumption is that users usually have a reservation price for a bundle, which refers to the maximum price that a customer is willing to pay for a bundle of products or services \cite{wu2019effective}. According to this assumption, users' purchase decisions are solely based on whether the price of the bundle is lower than their reservation price \cite{venkatesh1993probabilistic}. However, the simple assumption of reservation price fails to capture the complexity of users' actual purchasing behaviors, as other factors, such as the specific items included in the bundle and the discounts offered, can also impact their decision-making process \cite{harlam1995impact,giri2020bundle}. On the other hand, inter-temporal choice modeling focuses on studying the biases and factors that influence users' decision-making process. 
Existing works have primarily focused on studying projection bias and reference-point effect separately, and have not addressed situations where both biases occur simultaneously. Additionally, prior studies have primarily focused on developing models for these biases without providing algorithms or methods to learn users' personal biases from their transaction records. Furthermore, few works investigate these biases, specifically in the context of bundle sales and their implications for bundle strategies. In Section \ref{sec:related_work}, we provide a detailed overview of the existing works and their limitations in both bundle strategies and inter-temporal choice modeling. 

\textbf{Our Contributions.} In this study, we investigate inter-temporal choice modeling within the context of bundle sales. We propose the \textbf{PRO}jection \textbf{B}ias-\textbf{E}mbedded preference (\emph{Probe}) model to examine the cognitive biases present in users' purchase behaviors. Our approach is grounded in prospect theory from behavioral economics, which incorporates a weight function that captures users' projection bias in predicting the probability of needing additional items in the future, as well as a value function that models the reference-point effect in users' biased perceptions of price. The weight function includes a personalized parameter to account for the heterogeneity in users' projection biases and can accommodate both overestimation and underestimation scenarios. A novel method is proposed to learn their individual biases from historical records. Additionally, we present four methods to estimate users' expected prices within the value function. We provide a theoretical analysis to illustrate the impact of projection bias on the design of bundle sales strategies. Experimental results show that the proposed \emph{Probe} outperforms existing methods, and contributes to a better understanding of users' decisions when choosing between a bundle and a single item. The contributions of our work are as follows.
\begin{itemize}
    \item We investigate inter-temporal choice modeling in the context of bundle sales and propose \emph{Probe} to effectively model projection bias and the reference-point effect in users' decision-making processes.
    \item We theoretically analyze the impact of projection bias on the design of bundle pricing strategies.
    \item We propose a novel method to learn users' personal biases from historical records, and our experiments show that it achieves a better balance between precision and recall compared to prior works.
\end{itemize}


\section{Literature Review}
\label{sec:related_work}
Existing works related to our study can be categorized into two main topics: bundle strategies and inter-temporal choice modeling. In the following sections, we delve into each topic and discuss the methodologies employed in detail.

\subsection{Bundle Strategies}
Previous research on bundle strategies can generally be categorized into two main areas: bundle pricing and bundle recommendation. Bundle recommendation aims to  recommend a predefined bundle of items to individual users based on their preferences, while bundle pricing focuses on determining the optimal price for a bundle of products or services in order to maximize profits. 

\subsubsection{Bundle Recommendation}
In the bundle recommendation problem, techniques from recommendation systems are commonly leveraged to generate relevant bundle suggestions. For instance, Bayesian personalized ranking (BPR) is applied in \cite{pathak2017generating} to generate bundle recommendations. This approach involves generating item embeddings using the BPR method and subsequently generating bundle embeddings based on the embeddings of items included in the bundle. Another notable contribution in \cite{pathak2017generating} is the collection of a new dataset containing bundle information and user purchase behaviors from the Steam video game platform, which has been used in \cite{avny2022bruce,ma2022crosscbr}. 
In \cite{chang2021bundle}, a graph neural network is employed to learn the complex and high-order relationship among users, items, and bundles. This is achieved by unifying the user-item interaction, user-bundle interaction, and bundle-item affiliation into a heterogeneous graph, enabling an effective representation of these entities and their relationships. To address users' long-term interests, the work in \cite{he2022bundle} proposes a multi-round conversational recommendation framework, which incorporates a Markov decision process to model users' decision-making and feedback-handling processes in bundle contexts. Drawing inspiration from the sequence modeling capabilities of Transformers, the authors of \cite{avny2022bruce} adapt Transformers to generate representations of users, items, and bundles. This method aims to capture the latent relations among items within the bundle and understand users' preferences for individual items as well as the entire bundle. Furthermore, considering the user-bundle and user-item interactions corresponding to bundle views and item views, respectively, the work in \cite{ma2022crosscbr} models the cooperative association between the two different views through cross-view contrastive learning. This approach allows each view to distill complementary information from the other view, enhancing the self-discrimination of representations. 

However, the bundle recommendation problem is different from the inter-temporal choice problem studied in this work. The bundle recommendation problem typically focuses on scenarios where users do not have clear shopping needs and aim to discover new bundles that users may potentially be interested in. In contrast, our work considers a different scenario where users have specific shopping needs for an item and need to decide between purchasing the item at full price or opting for a bundle that includes the item at a discounted price. Moreover, typical approaches in the bundle recommendation problem concentrate on generating effective representations for users, items, and bundles. These representations are learned from users' historical purchase behavior. For example, if a user frequently purchases a particular item, the user's representation becomes closer to that of the item. Similarly, if an item is frequently included in bundles, the item's representation becomes close to that of the bundle. This approach implicitly encodes users' preferences within these representations. However, this approach lacks interpretability regarding why a user selects a specific item or a bundle and whether users' subjective biases exist in the selection process. In contrast, our work emphasizes explicit modeling of users' subjective biases that may exist in their decision-making processes, which may help us analyze and understand users' bundle choices.


\subsubsection{Bundle Pricing}

In the bundle pricing problem, a common assumption is that users usually have a reservation price for a bundle, which refers to the maximum price that a customer is willing to pay for a bundle of products or services \cite{venkatesh1993probabilistic}, and users' purchase decision is solely based on whether the actual price of the bundle is lower than their reservation price. Based on this assumption, prior works on bundle pricing schemes mainly focus on obtaining accurate estimations of consumers' reservation prices \cite{chen2022learning,ettl2020data,young2022mining}. In \cite{chen2022learning}, the authors propose a method to learn the distribution of users' reservation prices towards bundles using bundle sales data. They assume that customers' valuations of products follow a multivariate Gaussian distribution, and the valuation of a bundle is the sum of the valuations of its component products. The authors propose an algorithm that combines the expectation-maximization (EM) algorithm and Monte Carlo simulation to estimate the mean and the covariance matrix of the Gaussian distribution. The work in \cite{ettl2020data} jointly considers bundle recommendation and pricing, and proposes a personalized model that selects, prices, and recommends a bundle of related products to consumers during their online sessions. This model explicitly addresses the trade-off between myopic current profit and long-term profitability under inventory constraints. In contrast to the assumption of a monopoly in the prior works mentioned above, the study presented in \cite{young2022mining} introduces a more realistic scenario by considering the presence of a competitor firm selling the same inventory of items. The authors propose a scalable and efficient search heuristic to identify the optimal price that maximizes revenue in a market where only a monopoly offers the products, then the proposed heuristic is extended to address the situation in which a competitor is present in the market.

However, the simple assumption of reservation price \cite{venkatesh1993probabilistic} fails to capture the complexity of users' actual purchasing behaviors. In reality, users' choices are influenced by multiple factors in addition to price, e.g., the specific items included in the bundle and the discounts offered \cite{harlam1995impact,giri2020bundle}. Furthermore, users may possess biases in their perceptions of these factors, such as the projection bias \cite{loewenstein2003projection} and reference-point effect \cite{tversky1993context} discussed in Section \ref{sec:intro}. These biases can lead to suboptimal and irrational decisions. By overlooking the presence of biases, prior studies are limited in their ability to make accurate predictions and provide meaningful insights into users' bundle choices. 

\subsection{Inter-temporal Choice Modeling}
Inter-temporal choices refer to the decision-making process in which users make choices between options that have outcomes occurring at different points in time. 
Prior works on inter-temporal choice modeling mainly focus on the study of reward discounting, projection bias, and reference-point effect. Reward discounting refers to the tendency of users to assign less value or importance to future rewards compared to immediate rewards. Projection bias refers to the phenomenon that people tend to overestimate the similarity between their current preferences and future preferences. Reference-point effect refers to users' tendency to evaluate gains or losses from a decision relative to a reference point, rather than assessing them in absolute terms. In the following, we review prior works on these three topics.

\subsubsection{Reward Discounting}
Temporal reward discounting refers to the decrease in the subjective value of a delayed reward. The common discounting methods include hyperbolic discounting \cite{laibson1997golden} and exponential discounting \cite{mazur2009delay}. Hyperbolic discounting suggests that people tend to value immediate rewards more, and the discount rate over short horizons is relatively high while the discount rate over long horizons is relatively low.
On the other hand, exponential discounting assumes a constant rate of discount in the value of delayed rewards over time. Extensive research has examined the factors that influence reward discounts. For instance, in \cite{amasino2019amount}, the authors find that inter-temporal choice reflects the interaction of two distinct processes: one related to the discount of reward amounts and the other related to the discount of reward delays. The combination of these processes determines users' choices. Furthermore, the work in \cite{gershman2020rationally} proposes that users mentally simulate or imagine possible future outcomes, but these simulations are subject to noise or variability. The authors suggest that reward discounts can be explained by the optimal allocation of mental effort. When faced with larger rewards, users may allocate more mental effort to reduce the noise or uncertainty in their simulations, resulting in larger discounts. Recently, reward discounting has been used in reinforcement learning, where agents exhibit a preference for receiving rewards sooner rather than later \cite{schultheis2022reinforcement,zhou2021provably}. For instance, the work in \cite{schultheis2022reinforcement} presents a continuous-time model-based reinforcement learning method that can be generalized to arbitrary reward discount functions. However, in the context of bundle sales, the uncertainty of users' future demand for the additional item makes it challenging to determine the accurate delay in receiving the rewards. Consequently, we do not consider reward discounting in this work and will investigate this issue in our future work. 


\subsubsection{The Projection Bias}

One popular method specifically designed to model projection bias is the linear bias model \cite{loewenstein2003projection}. This approach assumes that the true values of users' current and future preferences are known. When users' choices are influenced by projection bias, the model assumes that users' estimates of future preferences are a linear combination of the true values of their current and future preferences. The linear bias model has been used to analyze users' irrational behaviors in various scenarios. For example, in the context of insurance purchase, they find that on days with severe air pollution, insurance sales increase by approximately $7.2\%$ \cite{chang2018something}.  In the context of fitness habits, researchers discovered that individuals with projection bias tend to underestimate the positive effects of fitness habits, leading to lower gym attendance \cite{acland2015naivete}. The work in \cite{kaufmann2022projection} examines the relationship between projection bias and work efficiency. They find that individuals with projection bias tend to underestimate the time required for the second task when transitioning from the first task. Although the linear bias model is popular and easy to analyze, it requires knowledge of the true values of users' future preferences, which may not be available. Furthermore, few works study how to estimate the parameters in this model, making it less suitable for quantitative analysis of users' irrational behaviors.

\subsubsection{The Reference-Point Effect}

To model the reference-point effect, one notable approach is prospect theory introduced by the Nobel Prize-winning economists Kahneman and Tversky \cite{kahneman1979prospect}. Prospect theory explains how individuals make decisions under uncertainty and incorporates the reference-point effect as a key component. According to this theory, individuals assess outcomes based on gains and losses relative to a reference point, and their preferences exhibit diminishing sensitivity to changes in outcomes as they move away from the reference point. One notable example is its utilization in the context of emergency decision-making. During emergencies, experts often encounter uncertain and complex situations, making it challenging to provide accurate decision-making information. In such scenarios, prospect theory is employed to capture the experts' bounded rationality and enhance the efficiency of emergency response \cite{ren2017hesitant}. Furthermore, prospect theory has been used to model and analyze the behavior of active consumers in response to real-time electricity pricing, and a prospect theory-based approach is proposed in \cite{jhala2018prospect} to study the interaction between an aggregator and multiple active consumers connected to a power distribution system. While prospect theory provides a valuable framework for modeling the reference-point effect, it does not address the issue of how individuals select the reference point. The selection of an appropriate reference point remains a challenge.


\section{The Proposed Preference Model}
\label{sec:framework}
\subsection{Problem Formulation}
\label{sec:formulation}
In this work, we consider an online shopping platform that contains $N$ individual items, and the platform can package some of these items into bundles to increase sales. Each item $i$ is characterized by two attributes: its price $c_i$ and its value in use $\xi_i$, and  we represent the item $i$ as a tuple $i=(c_i,\xi_i)$. 
The value in use $\xi_i$ denotes the utility that an item provides by satisfying human wants or serving a practical purpose. For example, an item with more features or of superior quality is perceived to have a higher value in use. Note that the value in use is an inherent characteristic of the item and is independent of whether users purchase or use the item \cite{macdonald2011assessing}.

\begin{figure*}[tbp]
\centering
  \centering
  \centerline{\includegraphics[width=13cm]{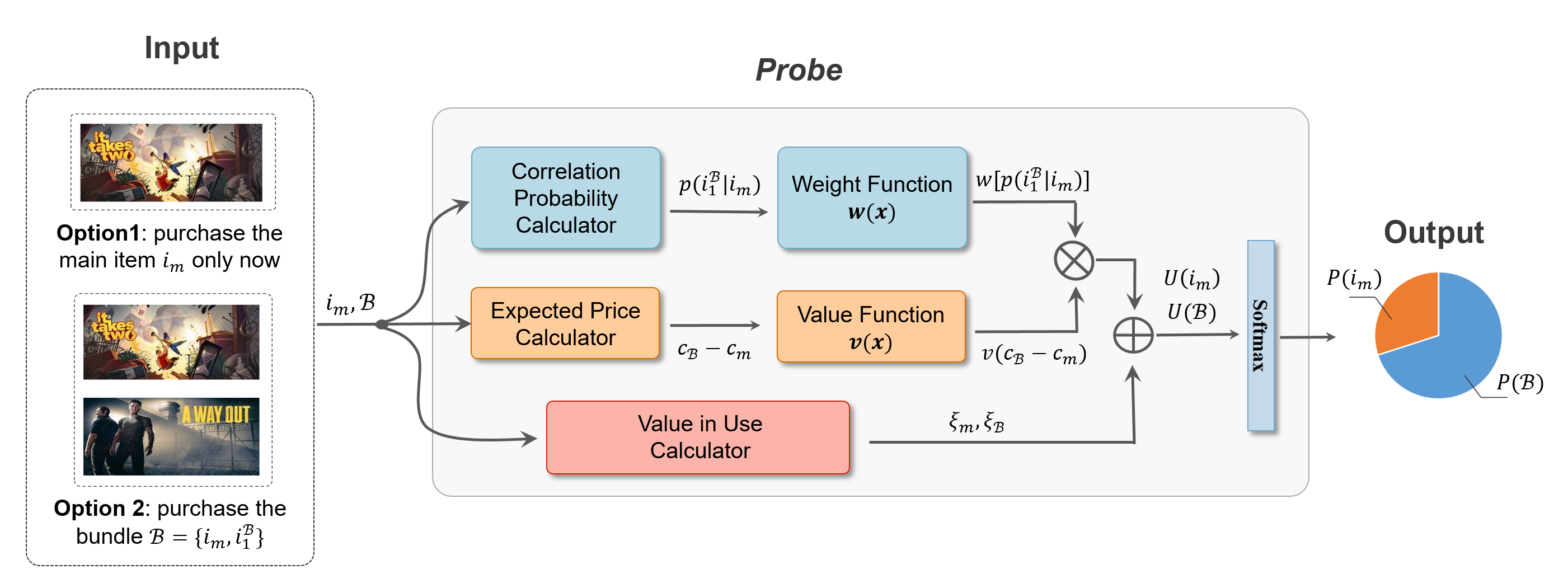}}
\caption{The proposed \emph{Probe} framework in a two-item bundle scenario.}
\label{fig:overall_framework}
\vspace{-0.4cm}
\end{figure*}

Consider a user creates a query for an item, and the platform reminds the user that a bundle containing this item is also on sale. We denote the individual item that users intend to purchase as $i_m$ and refer to it as the main item. Mathematically, we represent this item as $i_m = (c_m, \xi_m)$. Furthermore, we use $\mathcal{B}$ to denote the bundle, and $i^\mathcal{B}_1,i^\mathcal{B}_2,\cdots,i^\mathcal{B}_{|\mathcal{B}|-1}$ denote additional items included in the bundle. Thus, we have $\mathcal{B}=\{i_m,i^\mathcal{B}_1,i^\mathcal{B}_2,\cdots,i^\mathcal{B}_{|\mathcal{B}|-1}\}$, where $|\mathcal{B}|$ is the number of items in the bundle. The individual prices of additional items are $c^\mathcal{B}_1,c^\mathcal{B}_2,\cdots,c^\mathcal{B}_{|\mathcal{B}|-1}$, respectively, and the price of the bundle $\mathcal{B}$ is denoted as $c_{\mathcal{B}}$. We assume that $c_{\mathcal{B}} <c_m+c^\mathcal{B}_1+\cdots+c^\mathcal{B}_{|\mathcal{B}|-1}$, indicating that the price of the bundle is lower than the total price of these individual items. Additionally, we assume that $c_{\mathcal{B}}>c_m$, implying that the price of the bundle is higher than the price of the main item. Let $c_{\mathcal{B}}=r\left(\cdot \sum_{i_k\in \mathcal{B}} c_{i_k}\right)$, where $r$ represents the discount rate. As $r$ decreases, the price of the bundle decreases. Furthermore, we assume that the value of the bundle $\mathcal{B}$ is the sum of the values of all the items included in the bundle, i.e., $\xi_{\mathcal{B}} = \sum_{i_k \in \mathcal{B}} \xi_{i_k}$.

To facilitate the representation of bundles with varying numbers of items, we introduce a fixed-length vector $\boldsymbol{x}_{\mathcal{B}}\in \{0,1\}^{N}$. The length of $\boldsymbol{x}_{\mathcal{B}}$ is defined as $N$, and each element in the vector $\boldsymbol{x}_{\mathcal{B}}$ can only take the values of 0 or 1. 
If the $j$-th element in $\boldsymbol{x}_{\mathcal{B}}$ is 1, it indicates that the bundle $\mathcal{B}$ includes the $j$-th item. Otherwise, the $j$-th item is not in the bundle.

In this study, we propose \emph{Probe} to analyze users' cognitive biases within the bundle scenario. Fig. \ref{fig:overall_framework} illustrates the proposed Probe framework in a two-item bundle scenario. The input consists of two possible options of ``purchasing the main item $i_m$ only now'' and ``purchasing the bundle $\mathcal{B}=\{i_m,i_1^\mathcal{B}\}$''. Given the input, there are three parallel process flows.
\begin{itemize}
    \item The first process flow addresses users' projection bias. A correlation probability calculator is used to determine the actual probability that users will require the additional item $i_1^\mathcal{B}$ in the future, denoted as $p(i_1^\mathcal{B}|i_m)$. If there are more than 3 items in the bundle, we calculate this probability for all additional items as a whole. To model the projection bias, a weight function $w(x)$ is used to compute users' perceived probability.
    \item The second process flow addresses the reference-point effect. An expected price calculator is utilized to estimate users' expected price and compute the difference between the bundle price and the expected price. A value function $v(x)$ is used to evaluate users' perceived gains or losses based on this difference.
    \item The third process flow focuses on learning the value in use of the main item and the bundle, i.e., $\xi_m$ and $\xi_ {\mathcal{B}}$.
\end{itemize}
Next, we combine the outputs of the three process flows to obtain the utility when purchasing the main item only now $U(i_m)$, and the utility when purchasing the bundle $U(\mathcal{B})$. Then, the softmax function is applied to convert these utilities into the probability of selecting the main item $P(i_m)$ and the probability of selecting the bundle $P(\mathcal{B})$. That is, 
\begin{align}
\label{equ:prob}
  &P(i_m) = \frac{\exp\left[U(i_m)\right]}{\exp\left[U(i_m)\right]+\exp\left[U(\mathcal{B})\right]}, \quad \text{and}\\ \notag
  &P(\mathcal{B}) = \frac{\exp\left[U(\mathcal{B})\right]}{\exp\left[U(i_m)\right]+\exp\left[U(\mathcal{B})\right]} =1 - P(i_m).
\end{align}

In Fig. \ref{fig:overall_framework}, issues that need to be addressed include (Q1) how to construct the utility functions, i.e., $U(i_m)$ and $U(\mathcal{B})$, (Q2) how to model projection bias and reference-point effect, and (Q3) how to learn the value in use $\{ \xi_i \}$ and the unknown parameters in the three process flows. In the following, we first show our solutions to questions 1 and 2 in a two-item bundle scenario in Section \ref{sec:two_item_scenario}, then extend it to a multi-item bundle scenario in Section \ref{sec:multi_item_scenario}, and propose a learning algorithm in Section \ref{sec:estimation} to address question 3.

\vspace{-0.3cm}
\subsection{A Simple Two-Item Bundle Scenario}
\label{sec:two_item_scenario}
In this section, we use a two-item bundle, i.e., $\mathcal{B}=\{i_m,i_1^\mathcal{B}\}$, as an example to illustrate the proposed \emph{Probe} framework. In the following, we first introduce the proposed utility function. Next, we provide an introduction of projection bias modeling. Lastly, we show our model of the reference-point effect.

\subsubsection{Definition of the Utility Function} \label{sec:utilityfunc} Let $U(i_m)$ be the utility of purchasing the main item $i_m=(c_m, \xi_m)$ only now. For simplicity, we consider separable utility functions \cite{li2021prima++}, that is, 
\begin{align}
\label{equ:u_i_m}
U(i_m)=u_1(c_m)+u_2(\xi_m).
\end{align}
and our future work will investigate more complicated utility functions. Here, $u_1(c_m)$ and $u_2(\xi_m)$ are the utility values of paying $c_m$ and receiving the value in use $\xi_m$ now, respectively. Similarly, let $U(\mathcal{B})$ be the utility of purchasing the bundle $\mathcal{B}$, and we have 
\begin{align}
\label{equ:u_b}
U(\mathcal{B})=u_1(c_\mathcal{B})+u_2(\xi_\mathcal{B}).
\end{align}

To simplify the analysis, in this work, for the utility received from the value in use $u_2(\cdot)$, we employ a simple linear utility function. Specifically, we use $u_2(\xi_m) = \xi_m$ to represent the utility of the main item and $u_2(\xi_\mathcal{B}) = \xi_\mathcal{B}$ to represent the utility of the bundle. We assume that $u_2(\xi_\mathcal{B}) = \xi_\mathcal{B} = \xi_m + \xi_{i_1}$. It is worth mentioning that even with this simple linear utility function, simulations show that it gives superior performance when compared to previous approaches. We will investigate more complex functions for $u_2(\cdot)$ in our future work.

In the following, we focus on the analysis of $u_1(\cdot)$, the utility from the price. Note that there are three challenges that complicate the utility functions $u_1(c_m)$ and $u_1(c_\mathcal{B})$. 

\noindent \textbf{Addressing the Uncertainties in Users' Future Need for the Additional Item.} First, the price users will spend is a random variable due to the uncertainty associated with their future need for the additional item. 
\begin{itemize}
    \item If a user chooses to purchase the main item individually, there are two possibilities based on his/her future need for the additional item. When the user requires the additional item in the future, he/she will have to purchase it separately at its full price, resulting in a total price of $c_m+c_1$. On the other hand, when he/she does not need the additional item in the future, the price paid is that of the main item  $c_m$ only. 
    \item On the contrary, if the user purchases the bundle, the price they spend is fixed at $c_{\mathcal{B}}$, regardless of whether they will need the additional item in the future or not. 
\end{itemize}
Therefore, one critical issue in our bundle choice problems is estimating  users' actual future need for the additional item in the bundle. In this study, we address this issue by introducing the probability $p(i_1^\mathcal{B}|i_m)$, which 
represents the likelihood of users requiring item $i_1^\mathcal{B}$ after purchasing item $i_m$. The value of $p(i_1^\mathcal{B}|i_m)$ is influenced by the relationship between the main item $i_m$ and the additional item $i_1^\mathcal{B}$. For complementary items, such as a camera and a lens, it is probable that users will need the additional item in the future. In this case, $p(i_1^\mathcal{B}|i_m)$ will be close to 1. On the other hand, for substitute items that are seldom purchased together, such as two different brands of cameras, it is unlikely that users will require the additional item in the future. In this case, $p(i_1^\mathcal{B}|i_m)$ will be close to 0. We refer to $p(i_1^\mathcal{B}|i_m)$ as the \emph{correlation probability} since it reflects the correlation between the main item $i_m$ and the additional item $i_1^\mathcal{B}$. Additionally, we define $p(\bar i_1^\mathcal{B}|i_m)$ as the probability that users will not need the additional item $i_1^\mathcal{B}$ after purchasing $i_m$, and $p(\bar i_1^\mathcal{B}|i_m) = 1 - p(i_1^\mathcal{B}|i_m)$. In Section \ref{sec:estimation}, we provide details on how to estimate $p(i_1^\mathcal{B}|i_m)$ and $p(\bar i_1^\mathcal{B}|i_m)$ from users' previous purchasing records.

\noindent \textbf{Addressing the Reference-Point Effect.} \quad 
Furthermore, as mentioned in Section \ref{sec:intro}, users' satisfaction in purchasing the main item or bundle depends on the price difference relative to their expected price. The reference-point effect can introduce biases in users' price perceptions. For example, if the user's reference price is buying both the main item and the additional item at full price (that is, $c_m+c_1^\mathcal{B}$ in total), then purchasing the bundle at price $c_\mathcal{B}$ will give the user a gain of $c_m+c_1^\mathcal{B}-c_\mathcal{B}$. For different users, their reference prices will be different, and details of reference-point effect modeling is in Section \ref{sec:RefPointModel}.

\noindent \textbf{Addressing the Projection Bias.} \quad
From prospect theory, 
in our bundle choice problem, 
users' estimates of their potential gains/losses and the probability of their occurrence are often biased, and people make decisions based on their perceived gains/losses and perceived probability rather than the actual ones. To mathematically model users' perceived gains/losses and perceived probability, prospect theory employs a value function $v(x)$ and a weight function $w(p)$, respectively. 

\begin{figure}[tbp]
\centering
  \centering
  \centerline{\includegraphics[width=5cm]{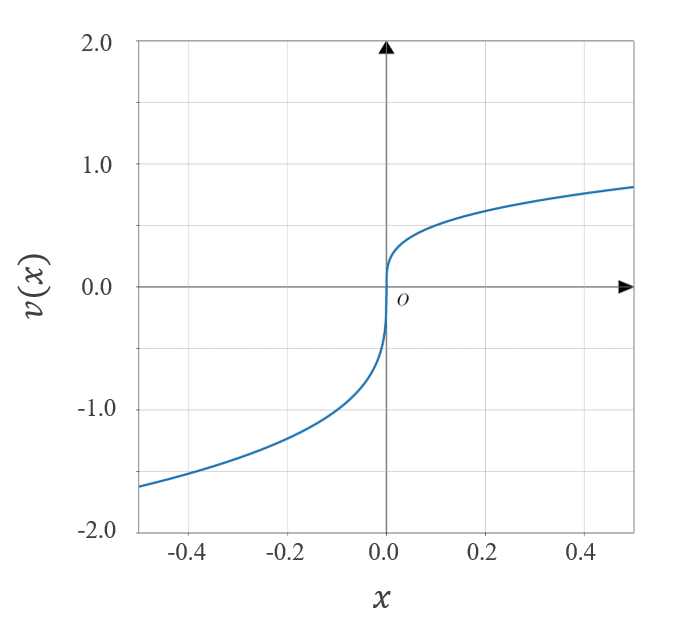}}
\caption{The value function in prospect theory.}
\label{fig:pt}
\vspace{-0.3cm}
\end{figure}

The value function $v(x)$ captures users' biases in perceiving gains and losses, and following the prior work in \cite{tversky1993context}, given $x$ as the difference between the actual price paid and the user's reference price, we let 
\begin{align}
\label{equ:pt_v}
  v(x) = \begin{cases}
           x^{\beta^+}, & \mbox{\text{if}\quad} x\geq 0, \\
           -\lambda(-x)^{\beta^-}, & \mbox{\text{if}\quad} x<0,
         \end{cases}
\end{align}
where 
$\beta^+,\beta^-\in (0,1)$ and $\lambda>1$ are hyper-parameters. Note that $v(0)=0$, that is, they perceive no loss and no gain at the reference point.
When $x > 0$, users perceive gains, and the value function takes the form of $x^{\beta^+}$. On the other hand, when $x < 0$, users perceive losses, and the value function becomes $-\lambda(-x)^{\beta^-}$. The value function captures the asymmetry in users' perceptions of gains and losses. 
Fig. \ref{fig:pt} shows an example of the value function with $\beta^+=\beta^-=0.3$ and $\lambda =2$. It exhibits two important characteristics: diminishing value and loss aversion. Diminishing value means that as the value of $x$ increases, the additional gains or losses brought to users by the same magnitude of increment $|\Delta x|$ are diminishing. Loss aversion means that 
users' fear of losses is greater than their joy of gains, and 
the negative utility perceived from the same amount of loss is greater than the positive utility perceived from the same amount of gain. This is reflected in the steeper slope of the value function for negative values of $x$. 

Given the actual probability $p$ that users may or may not need the additional item, the weight function captures the biases in users' perception of probabilities, and $w(p)$ is users' perceived probabilities. More discussions of the weight functions are in Section \ref{sec:ProjBias}.
Following the prior work in \cite{tversky1993context}, the weight function has distinct shapes for gains and losses. For gains with $x > 0$, the weight function is denoted as $w^+(p)$. For losses where $x < 0$, the weight function is denoted as $w^-(p)$.  

\subsubsection{Modeling the Perceived Gains/Losses} \label{sec:RefPointModel} 
A critical challenge in modeling the reference-point effect is to identify the expected price of users. From real-world bundle purchase experiences, we have observed that platforms typically emphasize the savings that can be obtained by selecting the bundle over the individual item. This approach is intended to make the bundle appear as a superior option and encourage users to select it. Drawing on this observation, we proposes that reference points can be established by considering users' different perceptions of thinking about savings or expenses of options.
In this work, we consider four different types of users, each corresponding to different modes of reasoning. \\
\textbf{Type I: Savings-centered users.} This type of user always selects the more expensive option as the reference point and estimates the potential gain (i.e., how much they save) by choosing each possible behavior. In our bundle choice problem, given they will need the additional item in the future, purchasing the bundle at the discounted rate costs $c_{\mathcal{B}}$, and purchasing the two items separately at full price costs $c_m+c_1^\mathcal{B}$. They select the more expensive option as the reference point, and thus, purchasing the bundle gives them an actual gain of $c_m+c_1^\mathcal{B}-c_{\mathcal{B}}$, i.e., the amount they save; while purchasing them separately gives them a gain of 0. Similarly, given they will not need the additional item in the future, they select the more expensive option of purchasing the bundle as the reference point. Therefore, purchasing the main item only saves them $c_{\mathcal{B}} - c_m$, while purchasing the bundle gives them a gain of 0. \\
\textbf{Type II: Expense-centered users.} This type of user is similar to the savings-centered users, while they select the less expensive option as the reference point and estimate the potential loss (i.e., how much more they need to pay) for each possible behavior. 
In our bundle choice problem, given they will need the additional item in the future, they select the less expensive option of purchasing the bundle as the reference point. Therefore, purchasing the two items separately will incur an actual loss of $c_m+c_1^\mathcal{B}-c_{\mathcal{B}}$, i.e., the extra money they need to pay; while purchasing the bundle gives them a loss of 0. Given they will not need the additional item in the future, purchasing the main item is the reference point. So purchasing the bundle will cost them $c_{\mathcal{B}}-c_m$ more; while purchasing the main item only incurs a loss of 0. \\
\textbf{Type III: Main item-centered users.} This type of user always uses the option of purchasing the main item now as the reference point and estimates the potential loss or gain of each possible behavior given they will need (or do not need) the additional item in the future. 
In our problem, given they will need the additional item in the future, the reference point is they purchase the main item now at price $c_m$ and purchase the additional item in the future at full price $c_1^\mathcal{B}$. Therefore, their actual gain is $c_m+c_1^\mathcal{B}-c_{\mathcal{B}}$ if they purchase the bundle; while if they purchase the main item now, they receive no loss and no gain. Given they will not need the additional item in the future, their reference point is purchasing the main item at price $c_m$. If they purchase the bundle, their actual loss is $c_{\mathcal{B}}-c_m$; while they perceive no gain and no loss if they purchase the main item only. \\
\textbf{Type IV: Bundle-centered users.} This type of user is similar to the Type III user, while they always use purchasing the bundle as the reference point. 
In our problem, given they will need the additional item in the future, the reference point is they purchase the bundle at price $c_{\mathcal{B}}$. Therefore, their actual loss is $c_m+c_1^\mathcal{B}-c_{\mathcal{B}}$ if they purchase the main item now; while if they purchase the bundle, they receive no loss and no gain. Given they will not need the additional item in the future, their reference point is purchasing the bundle at price $c_{\mathcal{B}}$. If they purchase the bundle, they receive no gain and no loss; while their actual gain is $c_{\mathcal{B}}-c_m$ if they purchase the main item only.

Given the above analysis of the reference point and users' actual gains/losses, we can use the value function $v(\cdot)$ in (\ref{equ:pt_v}) to find users' perceived gains/losses. 

\subsubsection{Modeling the Perceived Probabilities} \label{sec:ProjBias} 

Given $p(i_1^\mathcal{B}|i_m)$, the challenge in projection bias modeling is to select appropriate weight functions $w^+(p)$ and $w^-(p)$ to model bias in users' perception of probabilities. It is essential to highlight the distinction between projection bias and users' biased behaviors in the commonly studied gambling \cite{tversky1993context} and emergency decision-making scenarios \cite{ren2017hesitant}. 
These differences manifest in two key aspects. First, projection bias commonly arises in inter-temporal choices, where users face the challenge of weighing present costs against future benefits. In contrast, in gambling or emergency decision-making scenarios, users base their choices solely on immediate costs and benefits. Moreover, projection bias and users' behaviors in gambling or emergency decision-making scenarios exhibit distinct bias modes. According to studies \cite{kaufmann2022projection,loewenstein2003projection}, projection bias can lead to either overestimation or underestimation of users' future needs. For example, in our bundle choice problem, the occurrence of overestimation or underestimation depends on factors such as users' individual characteristics or the marketing strategy employed, and it is independent of the magnitude of the probability. In contrast, in gambling or emergency decision-making scenarios, the biases in users' estimates are related to the magnitude of the probability. This manifests as an overestimation of small probabilities and an underestimation of large probabilities.

\begin{figure}[tbp]
\centering
\begin{minipage}[b]{0.45\linewidth}
  \centering
  \centerline{\includegraphics[width=4.5cm]{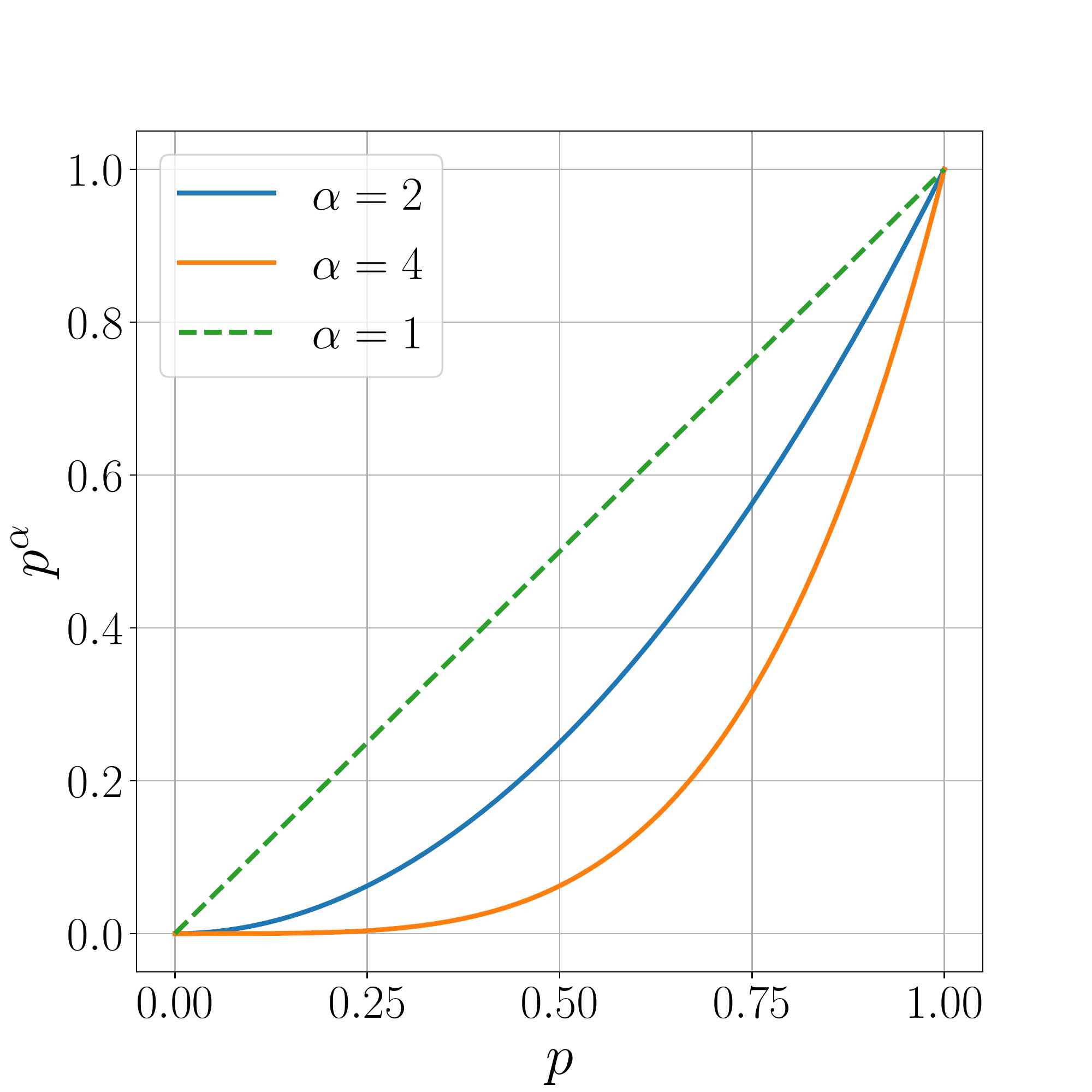}}
  \centerline{(a) The underestimation case}\medskip
\end{minipage}
\hfill
\begin{minipage}[b]{0.45\linewidth}
  \centering
  \centerline{\includegraphics[width=4.5cm]{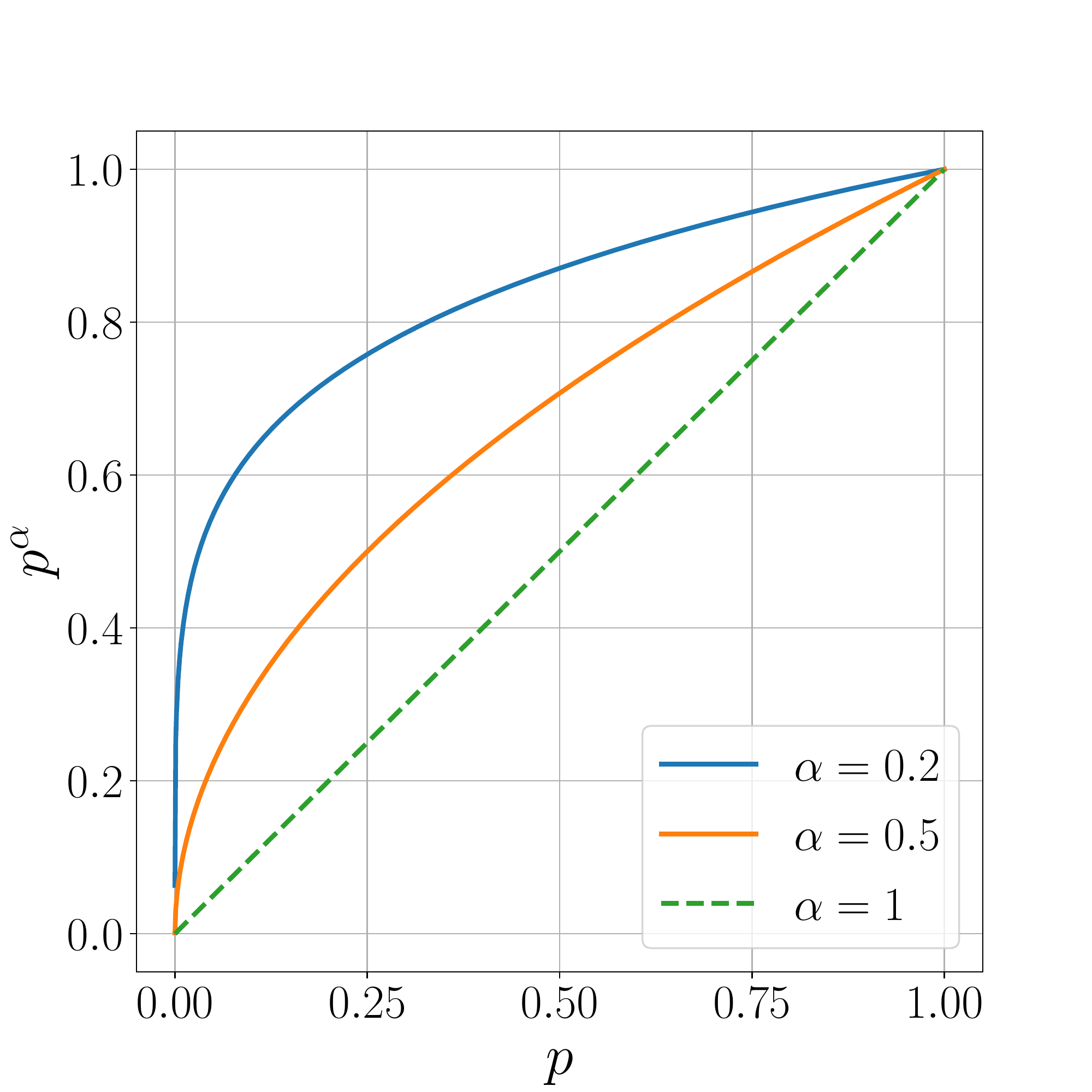}}
  \centerline{(b) The overestimation case }\medskip
\end{minipage}
\caption{The proposed weight function in the bundle choice scenario.}
\label{fig:w_alpha}
\vspace{-0.3cm}
\end{figure}

Taking into account the aforementioned distinctions, we propose a new weight function to capture users' perceived future preferences in the presence of projection bias for the bundle choice problem. Our proposed weight function is a power function, chosen for its simplicity and ability to differentiate between overestimation and underestimation scenarios. Specifically, the weight functions are defined as
\begin{equation}
\label{equ:w_p}
  w^+\left[p(i_1^\mathcal{B}|i_m)\right] = p(i_1^\mathcal{B}|i_m)^{\alpha^+}, \, w^-\left[p(\bar i_1^\mathcal{B}|i_m)\right]=p(\bar i_1^\mathcal{B}|i_m)^{\alpha^-}.
\end{equation}
Here, $w^+\left[p(i_1^\mathcal{B}|i_m)\right]$ and $p(i_1^\mathcal{B}|i_m)$ represent users' perceived probability and the true probability of users needing the additional item in the future, respectively. Similarly, $w^-\left[p(\bar i_1^\mathcal{B}|i_m)\right]$ and $p(\bar i_1^\mathcal{B}|i_m)$ represent users' perceived probability and the true probability of users not needing the additional item in the future. The parameters $\alpha^+$ and $\alpha^-$ denote the degree of deviation in the estimated future preference from the actual one and are referred to as the positive and negative bias coefficients, respectively. An example of the new weight function is shown in Fig.~\ref{fig:w_alpha}. Note that when $\alpha^+ = \alpha^- =1$, we have $w^+\left[p(i_1^\mathcal{B}|i_m)\right] = p(i_1^\mathcal{B}|i_m)$ and $w^-\left[p(\bar i_1^\mathcal{B}|i_m)\right]=p(\bar i_1^\mathcal{B}|i_m)$, and there is no projection bias where the perceived probabilities are the same as the actual ones. When $\alpha^+$ or $\alpha^-$ is greater than 1, it signifies an underestimation of $p(i_1^\mathcal{B}|i_m)$ in users' perceived probabilities, with a larger value indicating a greater bias. Conversely, when $\alpha^+$ or $\alpha^-$ is less than 1, it indicates an overestimation of $p(i_1^\mathcal{B}|i_m)$ in users' perceived probabilities, with a smaller value corresponding to a larger bias due to overestimation. 

Note that the sum of the weight function values may not equal 1, i.e., $w^+\left[p(i_1^\mathcal{B}|i_m)\right]+w^-\left[p(\bar i_1^\mathcal{B}|i_m)\right]\neq 1$. This deviation from strict probabilistic reasoning is a result of projection bias in users' perceived probabilities, similar to observations in prior works on prospect theory \cite{tversky1993context}. Furthermore, although we do not impose any restrictions on the relationship between $\alpha^+$ and $\alpha^-$ in our model, our experiment results in Section \ref{sec:quanlitativeanalysis} demonstrate an inverse relationship between $\alpha^+$ and $\alpha^-$, where a larger value of $\alpha^-$ often corresponds to a smaller value of $\alpha^+$. That is, when individuals overestimate $p(i_1^\mathcal{B}|i_m)$, they often underestimate $p(\bar i_1^\mathcal{B}|i_m)$. This observation aligns with human intuition \cite{ren2017hesitant}. 

Note that projection bias is not the same across users or items. To capture the variation in projection bias across users and items, we let 
\begin{equation}
\label{equ:alpha_p}
  \alpha^+ = (\alpha^+_u+\alpha^+_{i_m})/2, \quad \alpha^- = (\alpha_u^- + \alpha_{i_m}^-)/2.
\end{equation}
Here, $\alpha_u^+$ and $\alpha_u^-$ represent the impact of a user's characteristics on the biases, while $\alpha_{i_m}^+$ and $\alpha_{i_m}^-$ represent the impact of the main item on the biases. We refer to $\alpha_u^+$ and $\alpha_u^-$ as the user's positive and  negative bias coefficient, respectively, and refer to $\alpha_{i_m}^+$ and $\alpha_{i_m}^-$ as the item's positive and negative bias coefficient, respectively. The proposed weight function and bias coefficients offer a method to model users' perceived probabilities in a personalized manner.

\subsubsection{The Utility Functions and the Probabilities of Selecting the Main Item/Bundle}

Given the above analysis of the reference point selection and the projection bias, we can calculate the utility values $u_1(c_m)$ and $u_1(c_{\mathcal{B}})$ for different types of users in Section \ref{sec:RefPointModel}. 
Taking the example of savings-centered users, when they anticipate needing the additional item in the future, their perceived gains from purchasing the bundle are $v(c_m+c_1^\mathcal{B}-c_{\mathcal{B}})$, and their perceived gain/loss from purchasing the main item only now is 0 as $v(0)=0$. On the other hand, when these users do not expect to require the additional item in the future, their perceived gains from purchasing the main item is $i_m$ is $v(c_{\mathcal{B}}-c_m)$, and they perceive no gain and no loss if they purchase the bundle. By combining the value function with our proposed weight function, the utility values in terms of price are 
\begin{align}
\label{equ:u_t}
  u_1(c_m) &= w^-\left[p(\bar i_1^\mathcal{B}|i_m)\right]\cdot v(c_{\mathcal{B}}-c_m), \\ \notag u_1(c_\mathcal{B}) &= w^+\left[p(i_1^\mathcal{B}|i_m)\right] \cdot v(c_m+c_1^\mathcal{B}-c_{\mathcal{B}}).
\end{align}
Similarly, for expense-centered users, the utility values in terms of price are 
\begin{align}
u_1(c_m)&=w^+\left[p(i_1^\mathcal{B}|i_m)\right] \cdot v(c_{\mathcal{B}}-c_m-c_1^\mathcal{B}), \\ \notag
u_1(c_\mathcal{B})&= w^-\left[p(\bar i_1^\mathcal{B}|i_m)\right]\cdot v(c_m-c_{\mathcal{B}}).
\end{align}
For bundle-centered users, the utility values in terms of price are
\begin{align}
u_1(c_m)=0, \; u_1(c_\mathcal{B}) = w^+\left[p(i_1^\mathcal{B}|i_m)\right] \cdot v(c_m+c_1^\mathcal{B}-c_{\mathcal{B}}) \\ \notag 
+w^-\left[p(\bar i_1^\mathcal{B}|i_m)\right]\cdot v(c_m-c_{\mathcal{B}}).
\end{align}
For main item-centered users, the utility values in terms of price are
\begin{align}
u_1(c_\mathcal{B}) = 0, \; u_1(c_m)=w^+\left[p(i_1^\mathcal{B}|i_m)\right] \cdot v(c_{\mathcal{B}}-c_m-c_1^\mathcal{B}) \\ \notag 
+w^-\left[p(\bar i_1^\mathcal{B}|i_m)\right]\cdot v(c_{\mathcal{B}}-c_m).
\end{align}

Given the above $u_1(c_m)$ and $u_1(c_\mathcal{B})$, note that we let $u_2(\xi_m)=\xi_m$ and $u_2(\xi_\mathcal{B})=\xi_\mathcal{B}$ as stated in Section \ref{sec:two_item_scenario}, we can use Eq. (\ref{equ:u_i_m}) and Eq. (\ref{equ:u_b}) to obtain $U(i_m)$ and $U(\mathcal{B})$, respectively, and use Eq. (\ref{equ:prob}) to obtain $P(i_m)$ and $P(\mathcal{B})$.

\subsection{The Multi-Item Bundle Scenario}
\label{sec:multi_item_scenario}
In this section, we address the multi-item bundle scenario where $\mathcal{B}=\{i_m,i_1^\mathcal{B},i_2^\mathcal{B},\cdots,i_{|\mathcal{B}|-1}^\mathcal{B}\}$ with $|\mathcal{B}|>2$. We extend the proposed \emph{Probe} method from the two-item bundle scenario to handle this case. 
The utility functions in the multi-item bundle scenario remain the same as those in the two-item bundle scenario. Specifically, we calculate the utility value of the main item $i_m$ using Eq. \eqref{equ:u_i_m}, and the utility value of the bundle $\mathcal{B}$ using Eq. \eqref{equ:u_b}.

The challenge in the multi-item bundle scenario lies in dealing with users' future needs for multiple additional items. One intuitive approach is to discuss users' future needs for each additional item separately and list all possible combinations. However, the complexity of this approach grows exponentially, i.e. when the number of additional items is $||\mathcal{B}|-1$, there are $2^{|\mathcal{B}|-1}$ combinations to consider. Thus it is impractical for bundles with more than 10 items, which is common in scenarios such as game bundles or music bundles.

In this work, we consider the simple scenario where users choose between the following two options: either purchasing the entire bundle with all $|\mathcal{B}|$ items or purchasing the main item $i_m$ only. We ignore all other possible combinations of the items and plan to investigate effective algorithms to reduce the computation cost in our future work.
We introduce a virtual item $i_v^\mathcal{B}$ that represents the combined effect of all additional items in the bundle. The price of the virtual item $i_v^\mathcal{B}$ is set to be the sum of the prices of all additional items, denoted as $c_v^\mathcal{B}=\sum_{j=1}^{|\mathcal{B}|-1} c_j^\mathcal{B}$, and its value in use is the sum of the values in use of all additional items, denoted as $\xi_v^\mathcal{B}=\sum_{j=1}^{|\mathcal{B}|-1} \xi_j^\mathcal{B}$. 
In this case, we can simply replace the additional item $i_1^\mathcal{B}$ in the two-item bundle with the virtual item $i_v^\mathcal{B}$ in the multi-item scenario. This simplification allows us to apply the same framework and methodologies to model projection bias and the reference-point effect in both scenarios. 

Specifically, to model projection bias, $p(i_v^\mathcal{B}|i_m)$ is the probability that users will require the virtual item $i_v^\mathcal{B}$ after purchasing the main item $i_m$, and we have $p(\bar i_v^\mathcal{B}|i_m)=1-p(i_v^\mathcal{B}|i_m)$. The estimation of the probability $p(i_v^\mathcal{B}|i_m)$ is in Section \ref{sec:estimation}. We apply the weight function in Eq. \eqref{equ:w_p} to calculate users' perceived probabilities. Moreover, to model for the reference-point effect, we use the same four methods proposed in the two-item bundle scenario to determine users' expected price, where the price of the additional item $c_1^\mathcal{B}$ is replaced with the price of the virtual item $c_v^\mathcal{B}$. Subsequently, we apply the value function in Eq. \eqref{equ:pt_v} to calculate users' perceived gains or losses. For example, in the case of savings-centered users, the utility functions in the multi-item bundle scenario can be formulated as follows
\begin{align}
\label{equ:uti_multi_item}
  U(i_m) &= w^-\left[p(\bar i_v^\mathcal{B}|i_m)\right]\cdot v(c_{\mathcal{B}}-c_m)+\xi_{i_m}, \\ \notag
  U(\mathcal{B}) &= w^+\left[p(i_v^\mathcal{B}|i_m)\right] \cdot v(c_m+c_v^\mathcal{B}-c_{\mathcal{B}})+\xi_{\mathcal{B}}.
\end{align}
For other types of users, the analysis of $U(i_m)$ and $U(\mathcal{B})$ are the same and omitted.

\section{The Learning Algorithm}
\label{sec:estimation}

Given the above framework, we introduce our proposed learning algorithm to learn the parameters in the model. In our proposed \emph{Probe}, there are three sets of parameters that need to be learned: (1) the correlation probability $p(i_1^\mathcal{B}|i_m)$ in the two-item bundle scenario and $p(i_v^\mathcal{B}|i_m)$ in the multi-item bundle scenario. These probabilities represent the likelihood that users will require the additional item $i_1^\mathcal{B}$ or the virtual item $i_v^\mathcal{B}$ after purchasing the main item $i_m$; (2) the value in use of each item $\xi_{i}$; and (3) four bias coefficients, including the positive and negative bias coefficients of each user, i.e., $\alpha_u^+$ and $\alpha_u^-$, as well as the positive and negative bias coefficients of each item, i.e., $\alpha_{i}^+$ and $\alpha_{i}^-$. We utilize the example of savings-centered users to demonstrate the proposed learning algorithm. This method is also applicable to the other three types of users. To estimate these parameters, we first illustrate the estimation of the correlation probability and use $p(i_v^\mathcal{B}|i_m)$ as an example. The same method can also be used to estimate $p(i_1^\mathcal{B}|i_m)$. We then present our proposed learning algorithm to estimate the remaining parameters.

\subsection{Estimation of the Correlation Probability $p(i_v^\mathcal{B}|i_m)$}
\label{sec:correst} 

When estimating users' future need for the additional item $p(i_v^\mathcal{B}|i_m)$, there are two issues to address. First, different users have different needs for the additional item, and thus, $p(i_v^\mathcal{B}|i_m)$ will vary from person to person. However, in practical applications, the available information usually pertains to the items or bundles that users have purchased, while details such as users' income, consumption level, and individual characteristics often remain inaccessible. This makes it difficult to accurately estimate users' personal preferences for different items or bundles. In this work, we estimate and use the correlation probability averaged over the entire population in our analysis, and we will explore personal correlation probabilities in our future investigations.

Furthermore, the platform often needs to generate new bundles to fulfill users' dynamic interests. This requires that the estimation of $p(i_v^\mathcal{B}|i_m)$ be applied not only to the existing bundles in the training data but also to the new bundles that are not in the training data. Previous approaches, like the BPR method \cite{pathak2017generating}, BGCN \cite{chang2021bundle} and CrossCBR \cite{ma2022crosscbr}, typically learn vector representations for individual bundles and users, and leverage the similarity between bundles and users to estimate users' preferences for existing bundles. However, these methods have limitations in generating vector representations for new bundles and in predicting users' preferences for new bundles. 

To address this challenge, we assume that the estimation of the associated probability $p(i_v^\mathcal{B}|i_m)$ should depend on the individual items in the bundle and their relationships. 
The binary vector $\boldsymbol{x}_{\mathcal{B}}\in \{0,1\}^{N}$ represents the individual items contained in the bundle. 
Given $\boldsymbol{x}_{\mathcal{B}}$, we first mathematically model the relationship between $\boldsymbol{x}_{\mathcal{B}}$ and the associated probability $p(i_v^\mathcal{B}|i_m)$ and then propose a strategy to learn the unknown parameters 
from users' historical transaction records. Specifically, the estimation of the correlation probability $p(i_v^\mathcal{B}|i_m)$ involves two key issues: P1) how to mathematically model the relationship between the correlation probability $p(i_v^\mathcal{B}|i_m)$ and $\boldsymbol{x}_{\mathcal{ B}}$; and P2) how to learn the unknown parameters in the above model from users' past transaction records. These two key issues are discussed below.

\subsubsection{Modeling the relationship between $p(i_v^\mathcal{B}|i_m)$ and $\boldsymbol{x}_{\mathcal{ B}}$} 
\label{sec:formulation_p}

To study $p(i_v^\mathcal{B}|i_m)$, we initially define the ``co-purchase'' of item $i_j$ and $i_j$ as the situation where both items have been purchased by the same user. This includes cases where the user first purchases item $i_j$ and then purchases item $i_j$, as well as cases where the user purchases a bundle containing both $i_j$ and $i_k$. We compute the co-purchase matrix $\mathbf{F}\in\mathbb{R}^{N\times N}$ from users' previous shopping records. Here, $\mathbf{F}_{jk}$, with $j\neq k$, represents the number of co-purchases between the $j$-th and $k$-th items. As for diagonal elements, we set $\mathbf{F}_{jj}=0$ to avoid self-co-purchase counts.

It is important to note that the elements in matrix $\mathbf{F}$ exhibit a high degree of divergence. Specifically, some item combinations are frequently purchased together, leading to very large corresponding elements in $\mathbf{F}$, while other combinations are rarely purchased together, resulting in elements that are nearly zero. However, this high divergence of $\mathbf{F}$ may pose challenges when using the sigmoid function as the activation function during training. In particular, the gradient of the sigmoid function can approach zero when the input value is large, leading to the gradient vanishing problem. Conversely, when the input value is near zero, the gradient of the sigmoid function can be very large, resulting in the gradient exploding problem. These issues are also reported in \cite{yang2021rethinking}, which may prevent the model from converging to a desirable solution and may cause the optimization process to fail. To mitigate the aforementioned challenge, we adopt a standard technique known in graph theory to normalize the co-purchase matrix $\mathbf{F}$ \cite{merris1994laplacian}. Specifically, we normalize the matrix $\mathbf{F}$ as follows
\begin{equation}\label{equ:C_matrix}
  \mathbf{R} = \mathbf{D}^{-\frac{1}{2}}\mathbf{F}\mathbf{D}^{-\frac{1}{2}},
\end{equation}
where $\mathbf{D}$ is the degree matrix of $\mathbf{F}$. The matrix $\mathbf{D}$ is a diagonal matrix, where each element on the diagonal is the sum of all the elements in the corresponding row or column of $\mathbf{F}$, i.e., $\mathbf{D}_{jj} = \sum_k \mathbf{F}_{jk}$. The resulting matrix $\mathbf{R}$ is the normalized co-purchase matrix, where all elements in $\mathbf{R}$ are in the range $[0,1]$. Importantly, the normalized matrix $\mathbf{R}$ does not suffer from the gradient vanishing or gradients exploding problem, while the order relationship of the element values in $\mathbf{F}$ is preserved in $\mathbf{R}$.

Given the normalized co-purchase matrix $\mathbf{R}$, we model the relationship between the correlation probability $p(i_v^\mathcal{B}|i_m)$ and $\boldsymbol{x}_{\mathcal{ B}}$ as 
\begin{equation}\label{equ:esti_p}
  p(i_v^\mathcal{B}|i_m) = \sigma\left[\left(\boldsymbol{x}_{\mathcal{B}}\odot\mathbf{\Phi}_m\right)^T\cdot \mathbf{R}_m+b\right].
\end{equation}
Here, the vector $\mathbf{R}_m$ denotes the column in the normalized co-purchased matrix $\mathbf{R}$ that corresponds to the main item $i_m$. It is important to note that a higher frequency of co-purchases between two items usually indicates a higher correlation probability, but the degree of correlation between different item pairs may vary. For example, diapers and baby wipes are frequently purchased together and exhibit a substantial correlation. This phenomenon arises from their common usage in tandem during diaper changes for infants. On the other hand, for the example of diapers and beer, a classic case often cited in data analysis \cite{drucker2012practice}, although these items have been observed to be co-purchased frequently, they possess a notably lower correlation. One commonly suggested interpretation for the co-purchase behavior is that when men purchase diapers, they may reward themselves with a beer \cite{drucker2012practice}. In this work, we introduce a weight matrix $\mathbf\{\Phi\}\in\mathbb{R}^{N\times N}$ to model the varying strengths of this correlation between different item pairs, with larger weights indicating that the co-purchases of the two corresponding items contribute more to the estimation of the correlation probability than other item pairs. The vector $\mathbf{\Phi}_m$ is the column vector in the matrix $\mathbf{\Phi}$ that corresponds to the main item. The operation $\odot$ is the element-wise product (Hadamard product). In (\ref{equ:esti_p}), the term $\boldsymbol{x}_{\mathcal{B}}\odot\mathbf{\Phi}_m$ calculates the weights between the main item and the additional items in the bundle. The expression $\left(\boldsymbol{x}_{\mathcal{B}}\odot\mathbf{\Phi}_m\right)^T\cdot \mathbf{R}_m$ computes the weighted sum of the normalized co-purchase counts between the additional items and the main item. Furthermore, $b\in\mathbb{R}$ is the bias term, and the sigmoid function $\sigma(x)=1/\left[1+\exp(-x)\right]$ is used to confine the output value within the range $[0,1]$. 

\subsubsection{Learning unknown parameters in $p(i_v^\mathcal{B}|i_m)$}
In Eq. \eqref{equ:esti_p}, the parameters $\mathbf{\Phi}$ and $b$ play an important role in the correlation probability estimation, and thus it is critical to obtain an accurate estimation of these two parameters. To address this issue, we propose to formulate the estimation of $\mathbf{\Phi}$ and $b$ as a regression problem, i.e., estimating $\mathbf{\Phi}$ and $b$ by minimizing the difference between the estimated and the true values of $p(i_v^\mathcal{B}|i_m)$. 
In the training stage, 
we employ Ridge regression \cite{hoerl1970ridge}, a well-known linear regression technique, to optimize the parameters $\mathbf{\Phi}$ and $b$. Ridge regression aims to minimize the sum of squared errors between the predicted values and the target values, while additionally incorporating a penalty term that discourages the emergence of large parameter values. 

In this regression problem, 
the estimated value of the correlation probability $p(i_v^\mathcal{B}|i_m)$ can be computed using Eq. \eqref{equ:esti_p}, but the true value is difficult to obtain. To address this difficulty, we assume that the true value of $p(i_v^\mathcal{B}|i_m)$ can be approximated by the ratio of the number of joint purchases of the virtual item $i_v^\mathcal{B}$ and the main item $i_m$ to the number of purchases of the main item only $i_m$. In addition, we assume that the joint purchases of the virtual item $i_v^\mathcal{B}$ and the main item $i_m$ in the training data can be approximated by the purchases of the bundle ${\mathcal{B}}$. Specifically, assume that $M(i_m)$ and $M({\mathcal{B}})$ are the number of purchases of the main item $i_m$ and the bundle ${\mathcal{B}}$ in the training data, respectively. We use the ratio $M({\mathcal{B}})/M(i_m)$ to approximate the true value of $p(i_v^\mathcal{B}|i_m)$. 
We will investigate more accurate methods to determine the true value of $p(i_v^\mathcal{B}|i_m)$ in our future work. Note that the true value is only needed in the training stage, and in the testing stage, we compute the probability $p(i_v^\mathcal{B}|i_m)$  using Eq. \eqref{equ:esti_p}. 

\begin{algorithm}[tbp]
  \floatname{algorithm}{Algorithm}
  \renewcommand{\algorithmicrequire}{\textbf{input:}}
  \renewcommand{\algorithmicensure}{\textbf{output:}}
  \caption{The Proposed Learning Algorithm}
  \begin{algorithmic}[1]
  \Require The number of training data $K$, the number of iterations $T$, and $\mathbf{\Phi},b$,
  \Ensure $\alpha_u^+,\alpha_i^+,\alpha_u^-,\alpha_i^-$ and $\xi_{i}$
  \State initialize $\alpha_u^+,\alpha_i^+,\alpha_u^-,\alpha_i^-$ and $\xi_{i_m},\xi_{\mathcal{B}}$
  \For{$t=1,2,\cdots,T$}
    \State Shuffle the order of training data
    \For{$k=1,2,\cdots,K$}
        \State Compute $p(i_v^\mathcal{B}|i_m)$ according to Eq. \eqref{equ:esti_p}
        \State Compute $P({\mathcal{B}})$ and $P(i_m)$ according to Eq. \eqref{equ:prob}
        \State Compute gradients according to Eq. \eqref{equ:update}
        \State Update these parameters based on the computed gradients
    \EndFor
  \EndFor
  \end{algorithmic}
   \label{algo:training}
\end{algorithm}

\subsection{Learning Items' Value In Use and the Bias Coefficients} \label{sec:learningalg} 
Given the parameters $\mathbf{\Phi}$ and $b$, 
we propose the learning algorithm in Algorithm \ref{algo:training} to estimate the value in use of each item $\xi_{i}$, as well as the four bias coefficients, $\alpha_u^+,\alpha_u^-,\alpha_i^+,\alpha_i^-$. 


In our study, we employ the commonly used cross entropy function in the binary classification problem as our loss function. It quantifies the average discrepancy between the actual and predicted probability distributions, which is given by
\begin{equation}\label{equ:loss}
  L = -\frac{1}{K}\sum_{k=1}^K \left\{y^k_{\mathcal{B}}\log\left[P({\mathcal{B}})\right]
  +y^k_{i_m}\log\left[P(i_m)\right]\right\},
\end{equation}
where $K$ is the number of records in the training dataset. In Eq. (\ref{equ:loss}), in the $k$-th transaction record, $y^k_{\mathcal{B}}=1$ and $y^k_{i_m}=0$ when a user purchases the bundle, and $y^k_{\mathcal{B}}=0$ and $y^k_{i_m}=1$ when the user purchases the main item only. Additionally, due to a large number of samples in the training set, utilizing the gradient descent method will result in high computational overhead as it requires gradients for all samples. To accelerate the training speed, this work employs stochastic gradient descent \cite{bottou2012stochastic} as the optimization algorithm. Specifically, during each iteration, we randomly select a single data point, compute the gradients of the relevant parameters, and update them accordingly. We repeat this process over the training data for a total of $T$ times.

In the following, we use the parameter $\alpha_u^+$ and savings-centered users as an example to show the parameter update process. Similar update rules can be applied to other parameters and other types of users, and details can be found in the appendix. To update the value of $\alpha_u^+$ for savings-centered users, a single data point is randomly chosen, and the following update rule is employed
\begin{align}
\label{equ:update}
\alpha_u^+(t+1) & = \alpha_u^+(t) - \eta \cdot \nabla{\alpha_u^+}(t), \quad \mbox{where} \\ \notag  
\nabla{\alpha_u^+}(t) & = \frac{1}{2}\left[P^t({\mathcal{B}})-y_{\mathcal{ B}}\right]\cdot u^t_1(c_\mathcal{B})\cdot \ln\left[p(i_v^\mathcal{B}|i_m)\right].
\end{align}
In (\ref{equ:update}), $\alpha_u^+(t)$ and $\alpha_u^+(t+1)$ represent the values of $\alpha_u^+$ in the $t$-th and $(t+1)$-th iterations, respectively. $\nabla{\alpha_u^+}(t)$ is the gradient of $\alpha_u^+$ in the $t$-th iteration, and its derivation can be found in the supplementary file.
$\eta$ is the learning parameter that requires careful tuning to ensure effective convergence of the optimization process. $P^t({\mathcal{B}})$ is the predicted probability of selecting the bundle in the $t$-th iteration, and $u^t_1(c_\mathcal{B})$ is the computed utility of the bundle in the $t$-th iteration. The correlation probability $p(i_v^\mathcal{B}|i_m)$ is calculated using Eq. \eqref{equ:esti_p} with the optimized $\mathbf{\Phi}$ and $b$.

\subsection{The Testing Stage}

During the testing stage, given the learned parameters, when presented with two possible options, i.e., purchasing the main item $i_m$ only or purchasing the bundle $\mathcal{B}$, the following steps are carried out. 
First, given the normalized co-purchase matrix $\mathbf{R}_m$ and the parameters $\mathbf{\Phi}$ and $b$, we represent the new bundle $\mathcal{B}$ using its vector form $\boldsymbol{x}_{\mathcal{B}}$, and use \eqref{equ:esti_p} to predict the correlation probability $p(i_v^\mathcal{B}|i_m)$. Then, we use the utility functions in Section \ref{sec:two_item_scenario} to compute the utility values $U(i_m)$ and $U({\mathcal{B}})$ using the learned value in use $\xi_{i}$ and the four bias coefficients $\alpha_u^+$, $\alpha_u^-$, $\alpha_i^+$, and $\alpha_i^-$. Finally, we use \eqref{equ:prob} to compute the probabilities $P(i_m)$ and $P({\mathcal{B}})$, and the option with the highest probability is considered users' preferred choice.

\section{Theoretical Analysis of the Proposed \emph{Probe}}
\label{sec:analysis}

In this section, we theoretically analyze the impact of projection bias on the design of bundle sales strategies. We specifically focus on two important issues: (1) whether users with a higher projection bias are more inclined to purchase bundles; and (2) given the main item, how to select additional items to optimize the probability of users selecting the bundle. To facilitate our analysis, we use the two-item bundle scenario as an example, and we will explore the multi-item bundle scenario in the future. In this section, we consider savings-centered users as an example, and our analysis can be extended to other types of users.

\subsection{Positive Correlation Between Projection Bias and $P(\mathcal{B})$} 

Consider users who are faced with a choice between an item $i_m$ and a bundle $\mathcal{B}=\{i_m,i_1^\mathcal{B}\}$, where $c_{\mathcal{B}}-c_m>0$ and $c_m+c_1^\mathcal{B}-c_{\mathcal{B}}>0$. From \eqref{equ:w_p}, users with a higher projection bias tend to overestimate the probability that they will need the additional items (i.e., $p(i_1^\mathcal{B}|i_m)$) and underestimate the probability that they will not need the additional items (i.e., $p(\bar i_1^\mathcal{B}|i_m)$).

Following the discussion in Section \ref{sec:ProjBias}, we use $\alpha_u^+$ and $\alpha_u^-$ to quantify users' projection bias, as smaller values of $\alpha_u^+$ and larger values of $\alpha_u^-$ indicate a higher projection bias. We assume that $\alpha^+$ has a negative correlation with $\alpha^-$, which will be validated in Section \ref{sec:quanlitativeanalysis}. 
Specifically, according to Eq. \eqref{equ:w_p}, users with a higher projection bias tend to overestimate the probability that they will need the additional items (i.e., $p(i_1^\mathcal{B}|i_m)$) and underestimate the probability that they will not need the additional items (i.e., $p(\bar i_1^\mathcal{B}|i_m)$). 
In order to analyze the relationship between the probability that users will select the bundle $P(\mathcal{B})$ and the values of $\alpha_u^+$ and $\alpha_u^-$, we have Theorem \ref{theo:users}. The proof of Theorem \ref{theo:users} can be found in the appendix. Theorem \ref{theo:users} shows that $P(\mathcal{B})$ is a decreasing function of $\alpha_u^+$ and an increasing function of $\alpha_u^-$, and that users with a higher projection bias are more likely to purchase the bundle. The insights derived from Theorem \ref{theo:users} are consistent with the experimental observations outlined in Section \ref{sec:quanlitativeanalysis}. These findings provide valuable insights for the design of bundle sales strategies. For instance, platforms can estimate users' projection bias and target those with higher projection bias for effective bundle promotions.

\begin{theorem}
\label{theo:users}
Given an item $i_m$ and a bundle $\mathcal{B}=\{i_m,i_1^\mathcal{B}\}$ where $c_{\mathcal{B}}-c_m>0$ and $c_m+c_1^\mathcal{B}-c_{\mathcal{B}}>0$, we have
     \begin{align}
     \label{equ:dev_pb_alpha_plus}
        \frac{\partial P(\mathcal{B})}{\partial \alpha_u^+} 
        = \frac{1}{2}\cdot P(\mathcal{B}) P(i_m) u_1(c_\mathcal{B}) \cdot \ln \left[p(i_1^\mathcal{B}|i_m)\right]<0, \notag
    \end{align}
    \begin{align}
        \frac{\partial P(\mathcal{B})}{\partial \alpha_u^-} 
        = -\frac{1}{2}\cdot P(\mathcal{B}) P(i_m) u_1(c_{i_m}) \cdot \ln \left[p(\bar i_1^\mathcal{B}|i_m)\right]>0.
    \end{align}
That is, $P(\mathcal{B})$ is a decreasing function of $\alpha_u^+$, and is an increasing function of $\alpha_u^-$.
\end{theorem}


\subsection{Appropriate Additional Items for Bundle Sales}

In this section, given the main item, we study which additional item should be put in the bundle $\mathcal{B}$ to maximize the probability that users select the bundle. Specifically, we study the impact of the correlation probability $p(i_1^\mathcal{B}|i_m)$ and the price on the probability that users choose the bundle $P(\mathcal{B})$.

We first analyze the relationship between the probability $P(\mathcal{B})$ and the correlation probability $p(i_1^\mathcal{B}|i_m)$, and we have Theorem \ref{theo:correlation_probability} whose proof is in the appendix. Theorem \ref{theo:correlation_probability} emphasizes that, in order to increase the probability of users selecting the bundle $P(\mathcal{B})$, the platform should choose an additional item that has a high correlation probability with the main item $p(i_1^\mathcal{B}|i_m)$.

\begin{theorem}
\label{theo:correlation_probability}
Given the main item $i_m$ and a bundle $\mathcal{B}=\{i_m,i_1^\mathcal{B}\}$ where $c_{\mathcal{B}}-c_m>0$ and $c_m+c_1^\mathcal{B}-c_{\mathcal{B}}>0$, we have 
    \begin{small}
    \begin{align}
        \frac{\partial P(\mathcal{B})}{\partial p(i_1^\mathcal{B}|i_m)} 
            &= P(\mathcal{B}) P(i_m) \cdot
            \bigg\{\frac{\alpha_u^+ + \alpha_{i_m}^+}{2}\cdot u_1(c_\mathcal{B}) \cdot  \left[p(i_1^\mathcal{B}|i_m)\right]^{-1} \\ \notag
            & +\frac{\alpha_u^- + \alpha_{i_m}^-}{2}\cdot u_1(c_m) \cdot \left[1-p(i_1^\mathcal{B}|i_m)\right]^{-1}\bigg\}>0,
    \end{align}
    \end{small}
    and $P(\mathcal{B})$ is an increasing function of the correlation probability $p(i_1^\mathcal{B}|i_m)$.
\end{theorem}

Furthermore, we fix the bundle discount rate $r$ and analyze the relationship between the probability $P(\mathcal{B})$ and the price of the additional item $c_1^\mathcal{B}$, and our findings are in Theorem \ref{theo:price_additional_item}. 
\begin{theorem}
\label{theo:price_additional_item}
Consider the bundle $\mathcal{B}$ with a fixed discount rate $r$ that satisfies $r>{c_m}\big / {\sum_{i_k\in \mathcal{B}} c_{i_k}}$ (i.e., $c_{\mathcal{B}}-c_m>0$) and $r<1$ (i.e., $c_m+c_1^\mathcal{B}-c_{\mathcal{B}}>0$). Define
    \begin{small}
    \begin{align}
    \label{equ:def_A}
        \mathcal{A} &\coloneqq \sqrt[1-\beta^+]{\frac{\left[p(\bar i_1^\mathcal{B}|i_m)\right]^{(\alpha_u^- + \alpha_{i_m}^-)/2}}{\left[p(i_1^\mathcal{B}|i_m)\right]^{(\alpha_u^+ + \alpha_{i_m}^+)/2}}}>0, \quad 
        r_0 \coloneqq \frac{1}{1+\mathcal{A}^{\frac{1-\beta^+}{\beta^+}}}, \\ \notag
         &\text{and} \quad
        \kappa \coloneqq \frac{\left[(1-r)+\mathcal{A}\cdot \sqrt[1-\beta^+]{\frac{r}{(1-r)^{\beta^+}} }\right]}
        {r\cdot \left[1-\mathcal{A}\cdot \left(\frac{r}{1-r}\right)^{\frac{\beta^+}{1-\beta^+}}\right]}. 
    \end{align}
    \end{small}
We can then draw the following conclusions. 
    \begin{itemize}
        \item If $r<r_0$, then we have $\kappa>0$, and 
        \begin{align}
        \label{equ:low_rate}
            &\frac{\partial P(\mathcal{B})}{\partial c_1^\mathcal{B}}<0 \quad \text{when} \quad c_1^\mathcal{B} < \kappa\cdot c_m, \quad \text{and} \\ \notag
             &\frac{\partial P(\mathcal{B})}{\partial c_1^\mathcal{B}}>0 \quad \text{when} \quad c_1^\mathcal{B} > \kappa\cdot c_m.
        \end{align}
        \item If $r>r_0$, then we have $\kappa<0$, and 
        \begin{align}
        \label{equ:high_rate}
            \frac{\partial P(\mathcal{B})}{\partial c_1^\mathcal{B}}<0 \quad \text{when} \quad  c_1^\mathcal{B}\in (0,+\infty)
        \end{align} 
    \end{itemize}
\end{theorem}

From \eqref{equ:low_rate} and \eqref{equ:high_rate}, we observe that the partial derivative ${\partial P(\mathcal{B})}/{\partial c_1^\mathcal{B}}$ exhibits different behaviors depending on whether $r$ is below or above the threshold $r_0$. Specifically,  
\begin{itemize}
    \item when $r$ is below the threshold $r_0$, as $c_1^\mathcal{B}$ increases, the probability $P(\mathcal{B})$ first decreases and then increases, and $c_1^\mathcal{B}=\kappa\cdot c_m$ is the turning point. At this turning point, the probability of users purchasing the bundle is the lowest. That is, when the price of the additional item is significantly less than or greater than $\kappa\cdot c_m$, the probability of users purchasing the bundle is higher.
    \item On the other hand, when $r$ is above the threshold, the probability $P(\mathcal{B})$ is a decreasing function of $c_1^\mathcal{B}=\kappa\cdot c_m$. In other words, when the discount is small, a lower price for the additional item leads to a higher probability of users purchasing the bundle.
\end{itemize}

Moreover, we analyze the impact of projection bias, i.e., $\alpha_u^+$ and $\alpha_u^-$, on the threshold $r_0$ and the turning point $\kappa\cdot c_m$. To simplify the analysis, we assume that the item positive bias coefficient and item negative bias coefficient are equal to 0, i.e., $\alpha_{i_m}^+=\alpha_{i_m}^-=0$, and we observe the same trend for other values of $\alpha_{i_m}^+$ and $\alpha_{i_m}^-$.
\begin{itemize}
    \item First, we consider users without projection bias with $\alpha_u^-=\alpha_u^+=1$, and define $\left. \mathcal{A}^{wo} \coloneqq \mathcal{A} \right|_{\alpha_u^-=\alpha_u^+=1}$, $\left. r_0^{wo} \coloneqq r_0 \right|_{\alpha_u^-=\alpha_u^+=1}$ and $\left. \kappa^{wo} \coloneqq \mathcal{A} \right|_{\alpha_u^-=\alpha_u^+=1}$.
    \item Next, we consider users with projection bias with $\alpha_u^->1$ and $\alpha_u^+<1$. Define $\left. \mathcal{A}^{w} \coloneqq \mathcal{A} \right|_{\alpha_u^->1,\alpha_u^+<1}$, $\left. r_0^{w} \coloneqq r_0 \right|_{\alpha_u^->1, \alpha_u^+<1}$ and $\left. \kappa^{w} \coloneqq \mathcal{A} \right|_{\alpha_u^->1, \alpha_u^+<1}$.
\end{itemize} 
It is easy to show that $\mathcal{A}_w<\mathcal{A}_{wo}$, and $r_0^w>r_0^{wo}$. This implies that the threshold $r_0$ is higher when projection bias is taken into account. Furthermore, we have $\kappa_w<\kappa_{wo}$, indicating that the turning point is lower when projection bias is considered. This highlights the importance of accounting for projection bias in understanding the behavior of users and designing effective bundle sales strategies.


\section{Experiments}
\label{sec:test}

In this section, we first describe the experimental setups. Then, we present the performance comparisons and analyze the effects of parameter selections. Finally, we provide a qualitative analysis of users' projection bias.

\begin{table}[tbp]
  \centering
  \caption{The statistical description of the Steam dataset}
    \begin{tabular}{lr}
    \toprule
     \textbf{Items} & \textbf{Statistical values} \\
    \midrule
    $\#$ users & 20899 \\
    $\#$ individual games & 9136 \\
    $\#$ bundles & 229 \\
    $\#$ games contained in bundles & 684 \\
    $\#$ users who have never purchased a bundle & 10466 \\
    $\#$ users who have purchased at least one bundle & 10433 \\
    $\#$ individual game purchases & 159732 \\
    $\#$ bundle purchases & 37706 \\
    $\#$ purchase records & 197438 \\
    $\#$ purchase records per user & 9.45 \\
    \bottomrule
    \end{tabular}%
  \label{tab:dataset}%
\end{table}%

\subsection{Experimental Setup}
\label{sec:expsetup}
\textbf{Dataset Description.} 
We evaluate the performance of our proposed \emph{Probe} model using the Steam dataset \cite{pathak2017generating}. The Steam dataset is a comprehensive dataset that provides information on users' purchase and interaction behaviors with both individual games and bundles. The dataset is organized into three main parts: individual games data, bundles data, and users' interactions data. The individual game data provides detailed information about each game, including the game title, description, genre, price, and user reviews. The bundle data contains information including the bundle title, games included in the bundle, price, and the discount rate. The users' interaction data includes users' purchases, playtime, reviews, and ratings of individual games or bundles. 

Note that our access to the dataset is limited to users' final purchases, and we do not have explicit information about the available options users had when they made decisions. In particular, when we observe users buying a bundle, we do not know which specific item they originally intended to purchase. Similarly, when users' purchase involves a single item, we are unaware of the available bundles that were presented to them at the time of their decisions. To address this limitation, we use the following method to generate the available options. 
\begin{itemize}
    \item When there is a purchase of a bundle ${\mathcal{B}}$ in the dataset, we consider the game within the bundle that the user has played for the longest duration as the main item $i_m$. 
    \item When there is a purchase of an individual item $i_m$ in the dataset, we check bundles containing that specific item on the Steam platform. If there exists one bundle ${\mathcal{B}}$ that includes item $i_m$, we treat that bundle as the available option for the user. In cases where multiple bundles contain item $i_m$, we select the one with the lowest price. However, if there are no bundles containing item $i_m$, we discard that particular purchasing record of the single item as it does not provide relevant information for our analysis.
\end{itemize}
Following the above steps, we obtained a dataset with 197,438 records. Detailed statistical information is in Table~\ref{tab:dataset}.

\textbf{Evaluation Metrics.} In this work, we adopt the classic binary classification metrics to evaluate the performance of \emph{Probe}. Considering the sparsity of bundle purchases in the Steam dataset, where the number of bundle purchases (37,706) is approximately one-fifth of the number of individual game purchases (159,732), we define the positive class as bundle purchases and the negative class as individual game purchases. This allows us to account for the class imbalance in the dataset.  

\begin{table*}[tbp]
\centering
\caption{The confusion matrix}
\label{tab:confusion}
    \begin{tabular}{lcc}
    \toprule[1pt]
    \multicolumn{1}{p{18.58em}}{\diagbox{the real choice}{the predicted choice}} & purchase a bundle & purchase an individual game \\
    \midrule[0.6pt]
        purchase a bundle & TP & FN \\
        purchase an individual game & FP & TN \\
    \bottomrule[1pt]
    \end{tabular}%
\end{table*}

We use the confusion matrix shown in Table~\ref{tab:confusion} to calculate the precision, recall and F1 score, where
\begin{align}\label{equ:metric}
  &\text{precision}=\frac{TP}{TP+FP}, \quad \text{recall}=\frac{TP}{TP+FN}, \\ \notag
  & \mbox{and} \quad  \text{F1}=\frac{2\cdot \text{precision} \cdot \text{recall}}{\text{precision}+\text{recall}}.
\end{align}
Precision measures the accuracy of the model's positive predictions, and recall measures the model's ability to identify positive instances correctly. F1 score is the harmonic mean of precision and recall, providing a balanced measure between the two metrics. We primarily focus on F1 score as a comprehensive metric for evaluating the performance of the \emph{Probe}.

\textbf{Baseline Methods.} In this work, we compare the performance of our proposed \emph{Probe} with several baseline methods. Note that the linear bias model \cite{loewenstein2003projection}, which requires the true values of users' future preferences to be known, is not applicable in our setting. Therefore, we do not include the linear bias model in the baseline methods. 
\begin{itemize}
    \item Frequency-based method, where we assume that the probability of a user selecting the bundle in the testing set is the same as the frequency observed in the training set.
    \item Binary classification-based methods: Classic methods in this category include Naive Bayes Classifier \cite{taheri2013learning}, Support Vector Machines (SVM) \cite{chauhan2019problem}, Gradient Boosting Decision Trees (GBDT) \cite{ke2017lightgbm}, and AdaBoost \cite{ying2013advance}. Among these classifiers, AdaBoost achieves the best performance, thus, we show the results of AdaBoost as an example.
    \item Learning-to-rank methods including RankSVM \cite{lee2014large} and RankNet \cite{song2014adapting}.
    \item Bundle recommendation methods including BPR method \cite{pathak2017generating}, BGCN \cite{chang2021bundle}, and CrossCBR \cite{ma2022crosscbr}. 
\end{itemize}
Detailed implementation of these methods can be found in the appendix.

\begin{table}[tbp]
  \centering
  \caption{The experimental results}
    \begin{tabular}{cccc}
    \toprule
      Method & Precision & Recall & F1 \\
    \midrule
    Frequency-based method & 0.398 & 0.468 & 0.43 \\
    RankSVM \cite{lee2014large} & 0.236 & 0.867 & 0.371 \\
    BPR \cite{pathak2017generating} & 0.655 & 0.44 & 0.526 \\
    AdaBoost \cite{ying2013advance} & \textbf{0.748} & 0.593 & 0.661 \\
    RankNet \cite{song2014adapting} & 0.61 & 0.801 & \textbf{0.689} \\
    BGCN \cite{chang2021bundle} & 0.30 & \textbf{0.93} & 0.45 \\
    CrossCBR \cite{ma2022crosscbr}& 0.27 & 0.62 & 0.37 \\
    \hline
    Probe & 0.711 & 0.708 & \textbf{0.709} \\
    \bottomrule
    \end{tabular}%
  \label{tab:accuracy}%
\end{table}%

\subsection{Experimental results}
\label{sec:results}

\subsubsection{The Overall Performance}  \label{sec:overallperf}

The evaluation of the proposed \emph{Probe} method is performed using 5-fold cross-validation, and the process is repeated five times with different initial conditions. Furthermore, when considering all users as savings-centered users, it achieves the highest F1 score. Therefore, we employ this scenario as a benchmark in all our experiments, and more discussions of the reference-point model are presented in Section \ref{sec:perfrefpointmodel}. The average results are presented in Table~\ref{tab:accuracy}. Among the baseline methods, BGCN exhibits the highest recall but low precision, while AdaBoost achieves the highest precision but low recall. RankNet has the best performance in terms of the $F1$ score among all the baseline methods. Note that bundle recommendation methods such as BPR, BGCN, and CrossCBR solely rely on user-bundle and user-item interaction data, omitting crucial item and bundle features like prices. In contrast, RankNet and AdaBoost leverage these features, leading to their superior performance compared to the bundle recommendation methods. Impressively, the proposed \emph{Probe} exhibits a $2\%$ improvement in the $F1$ metric compared to RankNet, achieving the highest $F1$ score among all methods and effectively balancing precision and recall. Moreover, the proposed \emph{Probe} explicitly models users' decision-making processes within the bundle choice problem, endowing it with enhanced interpretability compared to previous works. Additionally, histograms of the predicted $P(i_m)$ and $P(B)$ can be found in the appendix.

\subsubsection{Effect of the Training Data Length} 
\label{sec:recordlength} 
To investigate the effect of the training data length on the performance, we select the top 4 methods with the highest $F1$ scores for performance comparison, namely, our proposed Probe, RankNet, AdaBoost, and BPR. 

In this experiment, we keep the testing set fixed with 39,488 records. Among the remaining $K=157,950$ records, we randomly select a part of them as the training dataset, and optimize the parameters using the learning algorithm in Section \ref{sec:learningalg}. Let $Q$ be the training set sampling rate. For example, when $Q=0.01$, only $1\%$ of the $157,950$ records are used as the training set.

\begin{figure}[tbp]
\centering
\centerline{\includegraphics[width=8cm]{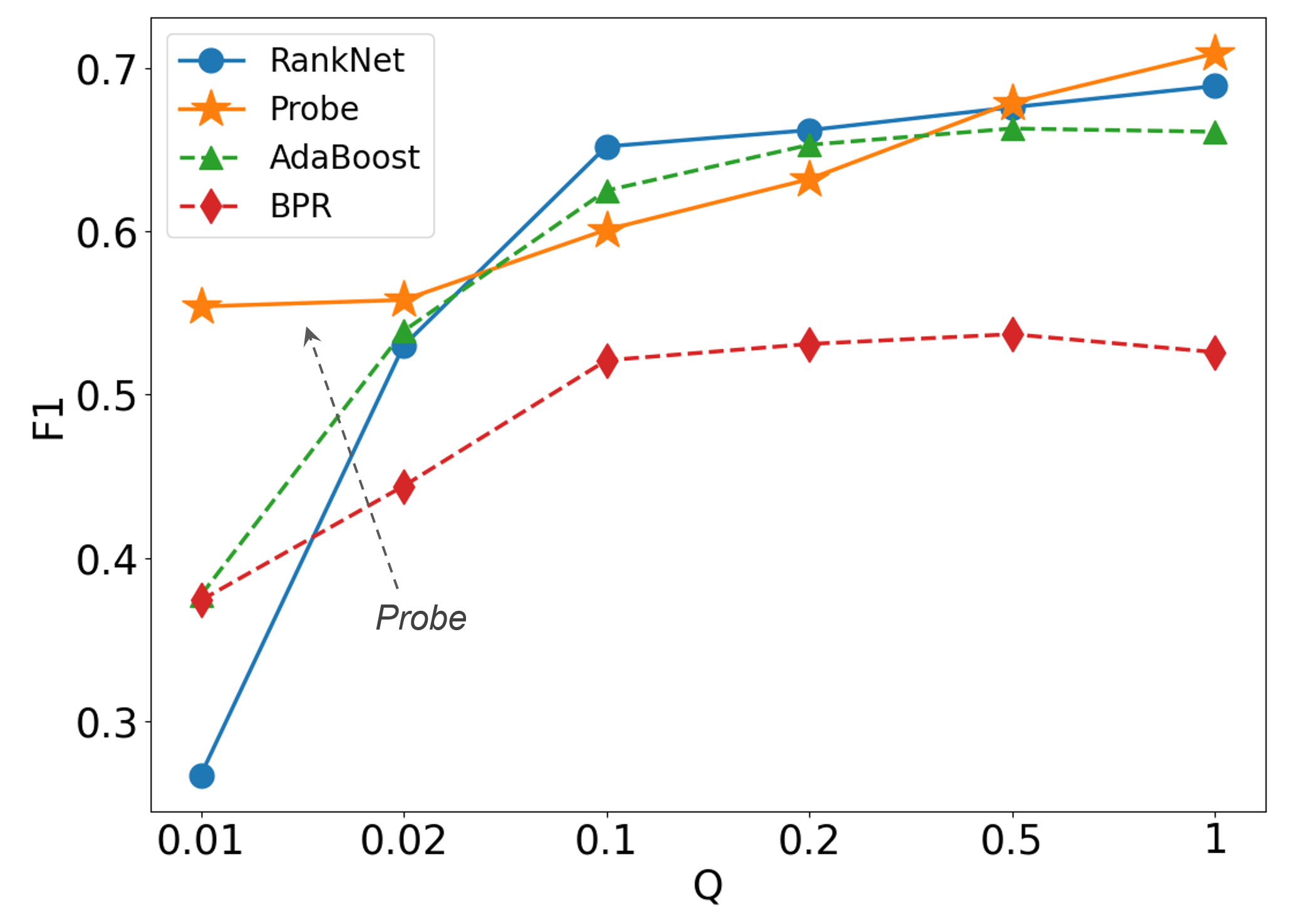}}
\caption{The effect of record length.}
\label{fig:size}
\end{figure}

The results are shown in Figure~\ref{fig:size}. Among all the baseline methods, RankNet initially has the worst F1 score at $Q=0.01$, and it exhibits the fastest improvement as $Q$ increases from 0.01 to 0.1. It achieves the second-best results when $Q=1$. The proposed \emph{Probe} method significantly outperforms the baseline methods at $Q=0.01$, and its performance increases steadily as more training data are available. It achieves the highest F1 score when $Q=1$. These results indicate that the proposed \emph{Probe} has a significant advantage when only a small amount of data is available and performs better than the baseline methods with larger datasets. 

\subsubsection{Evaluation of the Projection Bias Model} \label{sec:perfprojectionbiasmodel}

To validate our proposed personal projection bias model, we conduct experiments with different settings of $\alpha^+$ and $\alpha^-$. The results are shown in Table~\ref{tab:w_accu}. 

First, we investigate whether users have projection bias. We compare the performance of two sets of $\alpha^+$ and $\alpha^-$. In the first set labeled as ``W-I'' in Table~\ref{tab:w_accu}, we assume users have personal projection bias, and use  
\eqref{equ:alpha_p} to obtain personalized $\alpha^+$ and $\alpha^-$ for each user. In the second set named ``W-II'' in Table~\ref{tab:w_accu}, we assume users do not have projection bias and set $\alpha^+=\alpha^-=1$. 
Comparing the F1 scores of W-I and W-II, we find that W-I gives a higher $F1$ score, indicating that projection bias indeed exists in users' decision-making process.

Next, we study whether the projection bias is fixed for all users and considered two more scenarios labeled as ``W-III'' and ``W-IV'' in Table~\ref{tab:w_accu}. In W-III,  we assume that all users always overestimate their needs for items, and $\alpha^+$ and $\alpha^-$ are the same for all users. Similarly, in W-IV, we assume that all users always underestimate their needs for items, and $\alpha^+$ and $\alpha^-$ are the same for all users. Comparing the F1 scores of W-I, W-III, and W-IV, we find that W-I achieves the highest accuracy. These results indicate that 
users' projection bias varies from person to person.

Furthermore, we compare the performance of our proposed weight function (\ref{equ:w_p}) with the following weight functions 
\begin{align}
\label{equ:pt_w}
  &w^+(p)=\frac{p^{\gamma_1}}{\left[p^{\gamma_1}+(1-p)^{\gamma_1}\right]^{1/{\gamma_1}}}, \\ \notag
  & w^-(p)=\frac{p^{\gamma_2}}{\left[p^{\gamma_2}+(1-p)^{\gamma_2}\right]^{1/{\gamma_2}}},
\end{align}
which are commonly used in gambling \cite{tversky1993context} or emergency decision-making scenarios \cite{ren2017hesitant} and labeled as ``W-V'' in Table~\ref{tab:w_accu}. 
In (\ref{equ:pt_w}), $p$ is the actual probability of an outcome,  $\gamma_1$ and $\gamma_2$ are parameters that capture the degree to which users' perceptions are biased. 
The weight function in (\ref{equ:pt_w}) captures the tendency of users to overestimate small probabilities and underestimate large probabilities.
Comparing W-I and W-V, W-I achieves a higher F1 score, showing that our proposed weight function \eqref{equ:alpha_p} 
is more accurate in modeling users' projection bias in the bundle choice problem.

\begin{table}[tbp]
  \centering
  \caption{The effect of personalized projection bias}
    \begin{tabular}{lc}
    \toprule
    the weight function & F1 \\
    \midrule
    W-I: personal $\alpha^+,\alpha^-$ & \textbf{0.709} \\
    W-II: $\alpha^+=\alpha^-=1$ & 0.569 \\
    W-III: $\alpha^+=0.5,\alpha^-=2$ & 0.570 \\
    W-IV: $\alpha^+=2,\alpha^-=0.5$ & 0.570 \\
    W-V: $w^+(p)$ and $w^-(p)$ in Eq. \eqref{equ:pt_w} & 0.555 \\
    \bottomrule
    \end{tabular}%
  \label{tab:w_accu}%
\end{table}%

\subsubsection{Evaluation of the Reference-Point Effect Model}  \label{sec:perfrefpointmodel}
In this experiment, we assume that all users are of the same type when selecting the reference point, and Table \ref{tab:ref_accu} presents the results of the four different types of reference-point selections in Section \ref{sec:RefPointModel}. From Table \ref{tab:ref_accu}, when we assume all users are Type-I users, it achieves the highest $F1$ score. This indicates that most users are more concerned about the potential savings they can obtain from the available options. 
As a result, in this work, we assume that all users are savings-centered users who likely prioritize cost savings, and use this as an example in all experiments. In our future work, we plan to consider the more practical scenario where users have different types, study how to estimate their types and investigate the bundle choice problem with different types of users.

\begin{table}[tbp]
  \centering
  \caption{Performance of the four reference-point selection types}
    \begin{tabular}{cc}
    \toprule
    User Types & F1 \\
    \midrule
    Type I: Savings-centered users   & \textbf{0.709} \\
    Type II: Expense-centered users   &  0.553  \\
    Type III: Bundle-centered users  & 0.552  \\
    Type IV: Main item-centered users & 0.439\\
    \bottomrule
    \end{tabular}%
  \label{tab:ref_accu}%
\end{table}%

\begin{figure}[tbp]
\centering
\begin{minipage}[b]{0.45\linewidth}
  \centering
  \centerline{\includegraphics[width=4.3cm]{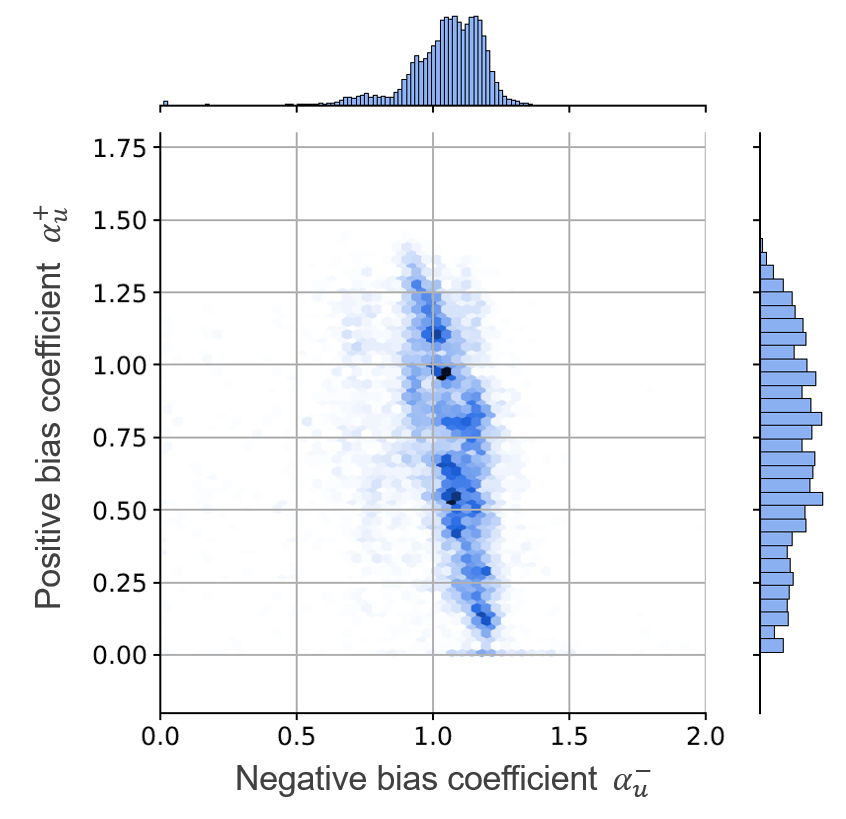}}
\end{minipage}
\hfill
\begin{minipage}[b]{0.45\linewidth}
  \centering
  \centerline{\includegraphics[width=4.3cm]{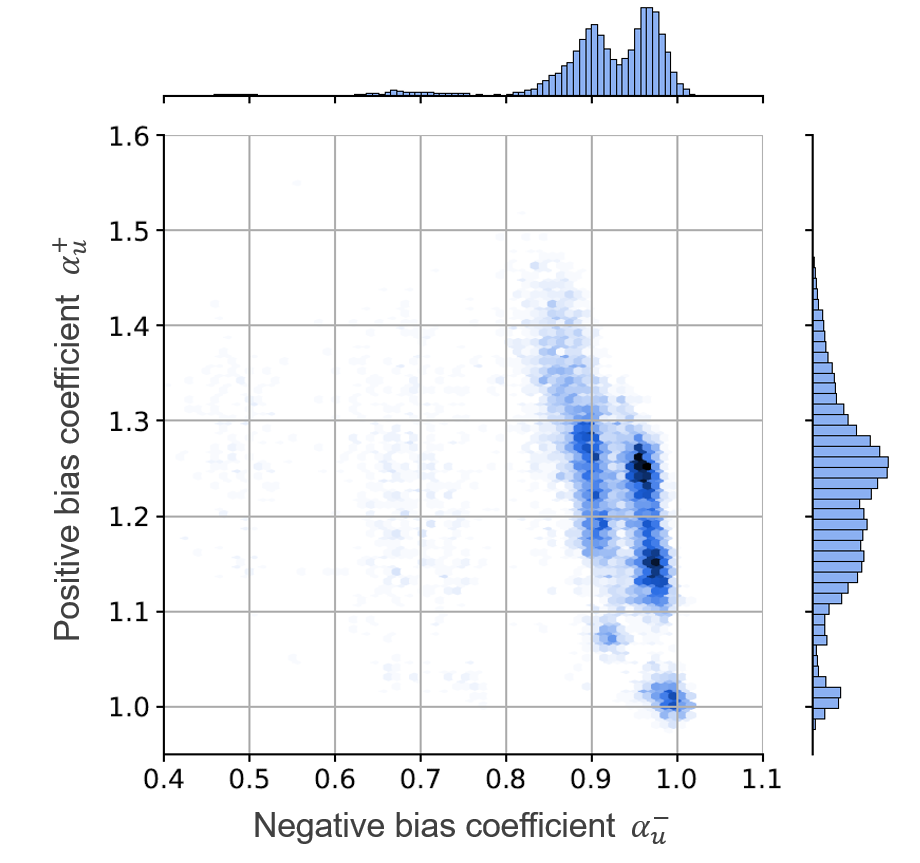}}
\end{minipage}
\caption{The joint distribution of two bias coefficients, where (a) users who have purchased at least a bundle, and (b) users who have never purchased a bundle. }
\label{fig:jointplot_alpha}
\end{figure}

\begin{figure}[tbp]
\centering
\begin{minipage}[b]{0.45\linewidth}
  \centering
  \centerline{\includegraphics[width=4.3cm]{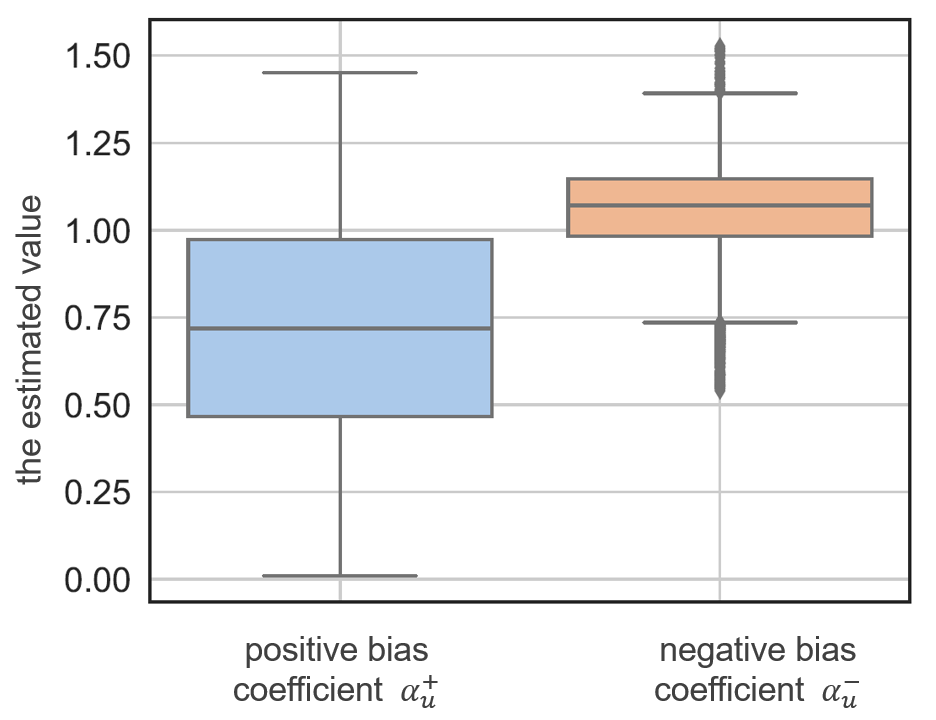}}
\end{minipage}
\hfill
\begin{minipage}[b]{0.45\linewidth}
  \centering
  \centerline{\includegraphics[width=4.3cm]{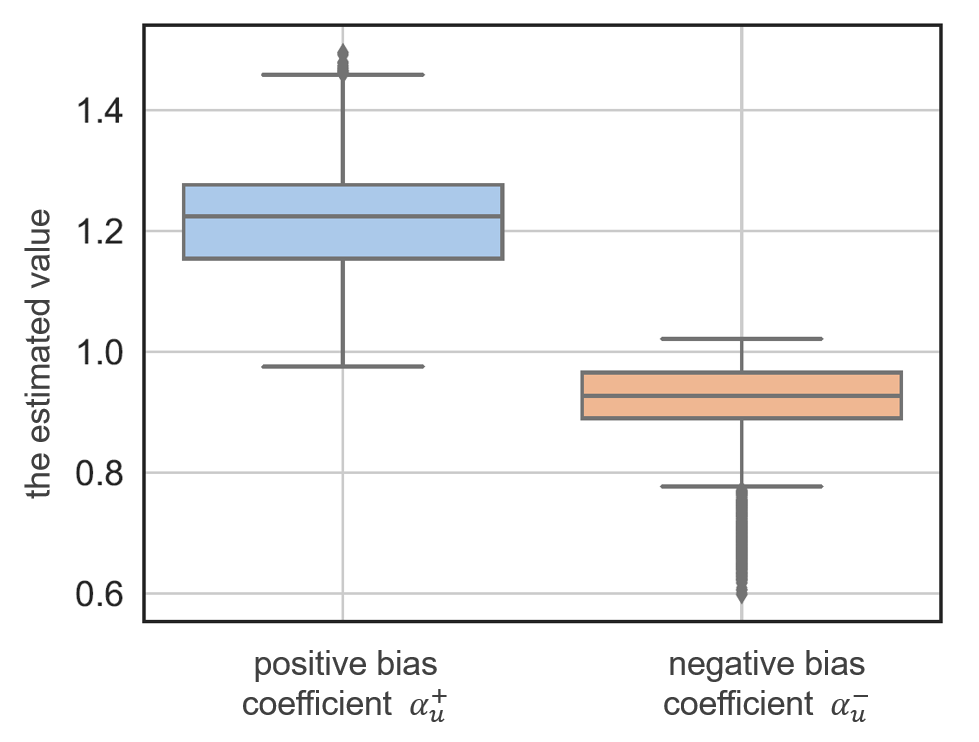}}
\end{minipage}
\caption{The box plots of two bias coefficients, where (a) users who have purchased at least a bundle, and (b) users who have never purchased a bundle.}
\label{fig:box_alpha}
\end{figure}

\begin{figure}[tbp]
\centering
\centerline{\includegraphics[width=6cm]{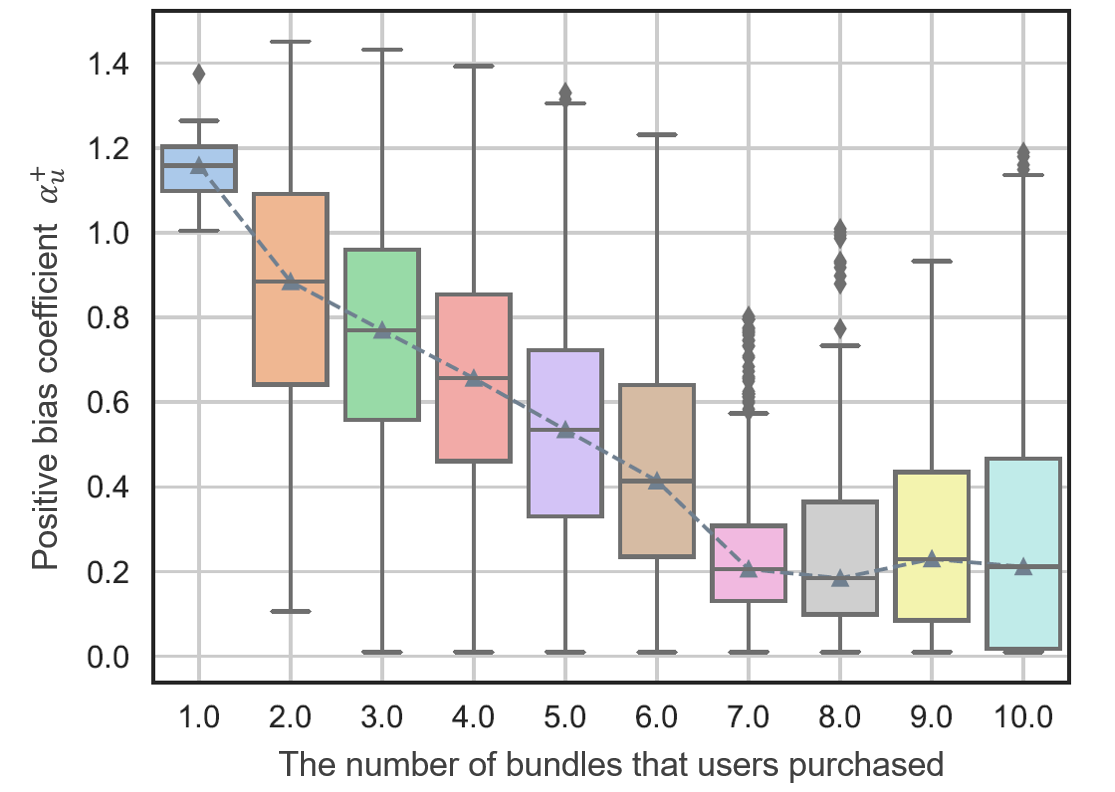}}
\caption{The relationship between the positive bias coefficient and the number of bundle purchases}
\label{fig:alpha_bundle_num}
\vspace{-0.5cm}
\end{figure}

\subsection{Qualitative Analysis of Users' Projection Bias} \label{sec:quanlitativeanalysis}
In this experiment, we aim to qualitatively investigate the relationship between the number of bundle purchases and users' projection bias. Users are categorized into two groups based on their purchase history: those who have purchased at least one bundle and those who have never purchased a bundle. We compare the positive bias coefficient $\alpha_u^+$ and the negative bias coefficient $\alpha_u^-$ for these two groups of users to study the relationship between users' projection bias and their bundle purchasing behavior.

Figure \ref{fig:jointplot_alpha} displays the joint distribution of the positive and negative bias coefficients for the two groups of users. The darker-colored areas indicate a higher number of users. In both groups of users, a negative correlation between the positive and negative bias coefficients is observed. This suggests that users are more likely to overestimate one probability while underestimating the other, rather than overestimating/underestimating both probabilities simultaneously.

Furthermore, Figure \ref{fig:box_alpha} presents the box plots of the positive and negative bias coefficients estimated using the proposed algorithm in Section \ref{sec:learningalg} for both groups of users. For users who have purchased at least one bundle, most of their positive bias coefficients are below 1.0, indicating a tendency to overestimate their demands for additional items. On the other hand, most of their negative bias coefficients are above 1.0, suggesting an inclination to underestimate the probability of not needing additional items. The overall distribution of positive bias coefficients is significantly lower than that of negative bias coefficients, reinforcing the finding that users with bundle purchases are more likely to overestimate their demands.

Similarly, for users who have never purchased a bundle, most of their positive bias coefficients are above 1.0, indicating a tendency to underestimate their demands for additional items. Conversely, most of their negative bias coefficients are below 1.0, suggesting an inclination to overestimate the probability of not needing additional items. The overall distribution of positive bias coefficients is significantly higher than that of negative bias coefficients, further supporting the observation that users without bundle purchases tend to underestimate their demands.

Additionally, for users who have purchased at least one bundle, we study the relationship between their positive bias coefficients and the number of bundle purchases. Figure \ref{fig:alpha_bundle_num} shows that the median of users' positive bias coefficients decreases as the number of bundle purchases increases. This indicates that users with a larger projection bias (and thus lower positive bias coefficients) are more likely to overestimate their demands for items and thus purchase more bundles.

\subsection{Summary}
In summary, when compared to prior works, the proposed \emph{Probe} method not only offers better interpretability but also outperforms prior works in terms of the F1 score, especially with limited training data. Additionally, our qualitative analysis reveals that users with a larger projection bias tend to overestimate their demands for items, leading to a higher likelihood of purchasing bundles. These findings contribute to a better understanding of users' decision-making processes in the bundle choice problem and highlight the effectiveness and interpretability of the proposed \emph{Probe}.

\section{Conclusion}
\label{sec:conclusion}
In this study, we explore users' intertemporal choices in the bundle sales scenario and focus on two prevalent biases: projection bias and the reference-point effect. We propose a preference model named \emph{Probe} to comprehensively analyze users' decision-making process. \emph{Probe} introduces prospect theory from behavioral economics to model users' choices, which includes a weight function that captures users' projection bias and a value function that captures reference-point effect. Based on \emph{Probe}, we propose a novel method for learning users' personal biases using historical records. Furthermore, we provide a comprehensive theoretical analysis on the impact of projection bias on the design of bundle sales strategies. Experimental results show that the proposed \emph{Probe} method outperforms prior works in terms of F1 score. Qualitative analysis reveals that users with a larger projection bias tend to overestimate their demand for items, leading to a higher likelihood of purchasing bundles. These findings contribute to a deeper understanding of users' decision-making mechanisms and can inform the design of optimal bundle strategies.

\bibliographystyle{IEEEtran}
\bibliography{refs}

\newpage

\section{Appendix}
\label{sec:appendix}

\subsection{The Proposed Learning Algorithm}
In the manuscript, a learning algorithm is proposed to estimate the remaining parameters, namely, the value in use of each item $\xi_{i}$, as well as the four bias coefficients, $\alpha_u^+,\alpha_u^-,\alpha_i^+,\alpha_i^-$. The gradients of these parameters are provided as follows.
    \begin{align}
    \label{equ:gradient}
        \nabla{\alpha_u^+}(t) &= \nabla{\alpha_i^+}(t) \\ \notag
        &= \frac{1}{2}\left[P^t({\mathcal{B}})-y_{\mathcal{ B}}\right]\cdot u^t_1(c_\mathcal{B})\cdot \ln\left[p(i_v^\mathcal{B}|i_m)\right], \\ \notag
        \nabla{\alpha_u^-}(t) &= \nabla{\alpha_i^-}(t) \\ \notag
        &= \frac{1}{2}\left[P^t(i_m)-y_{i_m}\right]\cdot u^t_1(c_m)\cdot \ln \left[p(\bar i_v^\mathcal{B}|i_m)\right], \\ \notag
        \nabla{\xi_{i_m}}(t) &= P^t(i_m)-y_{i_m}, \quad \mbox{and} \\ \notag
        \nabla{\xi_{i_k}}(t) &= P^t({\mathcal{B}})-y _{\mathcal{B}},\, \forall i_k\in \mathcal{B}, i_k\neq i_m.
    \end{align}

\begin{proof}
    For convenience, we drop the iteration index $t$ without confusion. Recall that given a single data point, the loss function is written as 
        \begin{equation}\label{equ:loss}
              L = -\left\{y_{\mathcal{B}}\log\left[P({\mathcal{B}})\right]
              +y_{i_m}\log\left[P(i_m)\right]\right\},
        \end{equation}
    where 
        \begin{align}
            \label{equ:prob}
              P(i_m) &= \frac{\exp\left[U(i_m)\right]}{\exp\left[U(i_m)\right]+\exp\left[U(\mathcal{B})\right]},
              \quad \text{and}  \\ \notag
              P(\mathcal{B}) &= \frac{\exp\left[U(\mathcal{B})\right]}{\exp\left[U(i_m)\right]+\exp\left[U(\mathcal{B})\right]}.
        \end{align}

    1) To begin with, we compute the gradient of $\alpha_u^+$ using the derivative rule. Since $U(\mathcal{B})$ is a function of $\alpha_u^+$, but $U(i_m)$ is not, the expression for the gradient is as follows
        \begin{align}
            \frac{\partial L}{\partial \alpha_u^+} 
            &= \frac{\partial L}{\partial U(\mathcal{B})}\cdot \frac{\partial U(\mathcal{B})}{\partial \alpha_u^+} \\ \notag
            &= \left[\frac{\partial L}{\partial P({\mathcal{B}})}\cdot \frac{\partial P({\mathcal{B}})}{\partial U(\mathcal{B})}+\frac{\partial L}{\partial P(i_m)}\cdot \frac{\partial P(i_m)}{\partial U(\mathcal{B})}\right]\cdot \frac{\partial U(\mathcal{B})}{\partial \alpha_u^+}.
        \end{align}
    In the following, we discuss how to compute each term in detail.\\
    1.1) Firstly, we have
        \begin{align}
        \label{equ:dev_L_PB}
            \frac{\partial L}{\partial P({\mathcal{B}})} = -y_{\mathcal{B}}\cdot\frac{1}{P({\mathcal{B}})},
        \end{align}
    and 
        \begin{align}
        \label{equ:dev_PB_UB}
            \frac{\partial P({\mathcal{B}})}{\partial U(\mathcal{B})} &= \frac{\exp\left[U(\mathcal{B})\right]\cdot \left\{\exp\left[U(i_m)\right]+\exp\left[U(\mathcal{B})\right]\right\}}{\left\{\exp\left[U(i_m)\right]+\exp\left[U(\mathcal{B})\right]\right\}^2} \\ \notag
            &\quad\quad  - \frac{\exp\left[U(\mathcal{B})\right]\cdot \exp\left[U(\mathcal{B})\right]}{\left\{\exp\left[U(i_m)\right]+\exp\left[U(\mathcal{B})\right]\right\}^2} \\ \notag
            & = \frac{\exp\left[U(\mathcal{B})\right]\exp\left[U(i_m)\right]}{\left\{\exp\left[U(i_m)\right]+\exp\left[U(\mathcal{B})\right]\right\}^2} \\ \notag
            & = P({\mathcal{B}}) P(i_m).
        \end{align}
    1.2) Secondly, we have
        \begin{align}
        \label{equ:dev_L_Pm}
            \frac{\partial L}{\partial P(i_m)} = -y_{i_m}\cdot\frac{1}{P(i_m)},
        \end{align}
    and 
        \begin{align}
        \label{equ:dev_Pm_UB}
            \frac{\partial P(i_m)}{\partial U(\mathcal{B})}  &= -\frac{\exp\left[U(\mathcal{B})\right]\exp\left[U(i_m)\right]}{\left\{\exp\left[U(i_m)\right]+\exp\left[U(\mathcal{B})\right]\right\}^2} \\ \notag
            &= -P({\mathcal{B}}) P(i_m).
        \end{align}
    Combining the results of the above two steps, we have
        \begin{align}
        \label{equ:dev_L_u_b}
            &\frac{\partial L}{\partial U(\mathcal{B})} 
            = \frac{\partial L}{\partial P({\mathcal{B}})}\cdot \frac{\partial P({\mathcal{B}})}{\partial U(\mathcal{B})}+\frac{\partial L}{\partial P(i_m)}\cdot \frac{\partial P(i_m)}{\partial U(\mathcal{B})} \\ \notag
            &= -y_{\mathcal{B}}\cdot\frac{1}{P({\mathcal{B}})}\cdot P({\mathcal{B}}) P(i_m) +y_{i_m}\cdot\frac{1}{P(i_m)}\cdot P({\mathcal{B}}) P(i_m) \\ \notag
            &= -y_{\mathcal{B}}\cdot P(i_m)+y_{i_m}P({\mathcal{B}}) \\ \notag
            &= -y_{\mathcal{B}}\cdot \left[1-P({\mathcal{B}})\right]+\left(1-y_{\mathcal{B}}\right)\cdot P({\mathcal{B}}) \\ \notag
            &= P({\mathcal{B}}) -y_{\mathcal{B}}.
        \end{align}
    1.3) Thirdly, we compute the derivatives of $U(\mathcal{B})$ with respect to $\alpha_u^+$. Notice that
        \begin{align}
            \frac{\partial U(\mathcal{B})}{\partial \alpha_u^+} = \frac{\partial \left[u_1(c_\mathcal{B})+u_2(\xi_{\mathcal{B}})\right]}{\partial \alpha_u^+} = \frac{\partial u_1(c_\mathcal{B})}{\partial \alpha_u^+},
        \end{align}
    and for saving-centered users, we have 
        \begin{align}
            u_1(c_\mathcal{B}) = \left[p(i_1^\mathcal{B}|i_m)\right]^{(\alpha_u^+ + \alpha_{i_m}^+)/2}  \cdot v(c_m+c_1^\mathcal{B}-c_{\mathcal{B}}).
        \end{align}
    Therefore, we have 
        \begin{align}
            \frac{\partial u_1(c_\mathcal{B})}{\partial \alpha_u^+} &= \frac{1}{2}\left\{\left[p(i_1^\mathcal{B}|i_m)\right]^{(\alpha_u^+ + \alpha_{i_m}^+)/2}\right\} \\ \notag
            &\qquad \cdot v(c_m+c_1^\mathcal{B}-c_{\mathcal{B}})\cdot \ln \left[p(i_1^\mathcal{B}|i_m)\right] \\ \notag
            &= \frac{1}{2} u_1(c_\mathcal{B}) \cdot \ln \left[p(i_1^\mathcal{B}|i_m)\right],
        \end{align}
    and, 
        \begin{align}
        \label{equ:dev_ub_alpha_plus}
            \frac{\partial U(\mathcal{B})}{\partial \alpha_u^+} = \frac{1}{2} u_1(c_\mathcal{B}) \cdot \ln \left[p(i_1^\mathcal{B}|i_m)\right].
        \end{align}
   Combining the results in Eq. \eqref{equ:dev_L_u_b} and \eqref{equ:dev_ub_alpha_plus}, we have 
        \begin{align}
            \frac{\partial L}{\partial \alpha_u^+} 
            &= \frac{\partial L}{\partial U(\mathcal{B})}\cdot \frac{\partial U(\mathcal{B})}{\partial \alpha_u^+} \\ \notag
            &= \frac{1}{2}\left[P({\mathcal{B}})-y_{\mathcal{ B}}\right]\cdot u_1(c_\mathcal{B})\cdot \ln\left[p(i_v^\mathcal{B}|i_m)\right]
        \end{align}

    2) The derivative of the loss function $L$ with respect to $\alpha_i^+$ is the same as the derivative of $L$ with respect to $\alpha_u^+$, and is therefore omitted in this paper. 

    3) Next, we compute the derivative of the loss function $L$ with respect to $\alpha_u^-$. Since $U(i_m)$ is a function of $\alpha_u^-$, but $U(\mathcal{B})$ is not, the expression for the gradient is as follows
        \begin{align}
            \frac{\partial L}{\partial \alpha_u^-} 
            &= \frac{\partial L}{\partial U(i_m)}\cdot \frac{\partial U(i_m)}{\partial \alpha_u^+} \\ \notag
            &= \left[\frac{\partial L}{\partial P({\mathcal{B}})}\cdot \frac{\partial P({\mathcal{B}})}{\partial U(i_m)}+\frac{\partial L}{\partial P(i_m)}\cdot \frac{\partial P(i_m)}{\partial U(i_m)}\right]\cdot \frac{\partial U(i_m)}{\partial \alpha_u^-}.
        \end{align}
    3.1) Similar to Eq. \eqref{equ:dev_PB_UB} and \eqref{equ:dev_Pm_UB}, we have 
        \begin{align}
        \label{equ:dev_pb_um}
            \frac{\partial P({\mathcal{B}})}{\partial U(i_m)} =  -P({\mathcal{B}}) P(i_m),
        \end{align} 
        and 
        \begin{align}
        \label{equ:dev_pm_um}
            \frac{\partial P(i_m)}{\partial U(i_m)} = P({\mathcal{B}}) P(i_m).
        \end{align}
    Combining the results of Eq. \eqref{equ:dev_L_PB}, \eqref{equ:dev_L_Pm}, \eqref{equ:dev_pb_um} and \eqref{equ:dev_pm_um}, we have 
        \begin{align}
        \label{equ:dev_L_um}
            \frac{\partial L}{\partial U(i_m)} = P(i_m)-y_{i_m}
        \end{align}
    3.2) Further, we have 
        \begin{align}
            \frac{\partial U(i_m)}{\partial \alpha_u^-} = \frac{\partial \left[u_1(c_m)+u_2(\xi_{i_m})\right]}{\partial \alpha_u^-} = \frac{\partial u_1(c_m)}{\partial \alpha_u^-}.
        \end{align}
    Recall that 
        \begin{align}
            u_1(c_m) = \left[p(\bar i_1^\mathcal{B}|i_m)\right]^{(\alpha_u^- + \alpha_{i_m}^-)/2}\cdot v(c_{\mathcal{B}}-c_m),
        \end{align}
    then we have
        \begin{align}
        \label{equ:dev_um_alpha_minus}
            \frac{\partial u_1(c_m)}{\partial \alpha_u^-} = \frac{1}{2}\cdot u_1(c_{i_m}) \cdot \ln \left[p(\bar i_1^\mathcal{B}|i_m)\right].
        \end{align}
    Combining the results of Eq. \eqref{equ:dev_L_um} and Eq. \eqref{equ:dev_um_alpha_minus}, we have 
        \begin{align}
            \frac{\partial L}{\partial \alpha_u^-} = \frac{1}{2}\left[P(i_m)-y_{i_m}\right]\cdot u_1(c_m)\cdot \ln \left[p(\bar i_v^\mathcal{B}|i_m)\right].
        \end{align}

    4) The derivative of the loss function $L$ with respect to $\alpha_i^-$ is the same as the derivative of $L$ with respect to $\alpha_u^-$, and is therefore omitted in this paper.

    5) At last, we compute the derivative of the loss function $L$ with respect to the value in use for each item $\xi_{i}$. For the main item, we have 
        \begin{align}
            \frac{\partial L}{\partial \xi_{i_m}} 
                = \frac{\partial L}{\partial U(i_m)}\cdot \frac{\partial U(i_m)}{\partial \xi_{i_m}}.
        \end{align}
    Since 
        \begin{align}
            \frac{\partial U(i_m)}{\partial \xi_{i_m}} = \frac{\partial \left[u_1(c_m)+u_2(\xi_{i_m})\right]}{\partial \xi_{i_m}} = \frac{\partial u_2(\xi_{i_m})}{\partial \xi_{i_m}}=1,
        \end{align}
    we have 
        \begin{align}
            \frac{\partial L}{\partial \xi_{i_m}} = \frac{\partial L}{\partial U(i_m)} = P(i_m)-y_{i_m}.
        \end{align}
    For additional items $i_k$ in the bundle with $i_k\neq i_m$, following the similar analysis, we have 
        \begin{align}
            \frac{\partial L}{\partial \xi_{\mathcal{B}}} &= \frac{\partial L}{\partial U(\mathcal{B})} = P({\mathcal{B}})-y _{\mathcal{B}}, \quad \text{and} \\ \notag
            \frac{\partial \xi_{\mathcal{B}}}{\partial \xi_{i_k}} &= \frac{\partial \sum_{i_k \in \mathcal{B}} \xi_{i_k}}{\partial \xi_{i_k}} = 1.
        \end{align}
    Therefore, we have  
        \begin{align}
            \frac{\partial L}{\partial \xi_{i_k}} = P({\mathcal{B}})-y _{\mathcal{B}},
        \end{align}

\end{proof}

\subsection{Theoretical Analysis of the Proposed \emph{Probe}}


\textbf{Proof of Theorem 1.} The proof of Theorem 1 in the manuscript is as follows.
\begin{proof}
    Recall that the probability that users choose the bundle is formulated as 
    \begin{align}
     P(\mathcal{B})= \frac{\exp\left[U(\mathcal{B})\right]}{\exp\left[U(i_m)\right]+\exp\left[U(\mathcal{B})\right]}. \notag
    \end{align}
    
    1) We first consider the derivative of $P(\mathcal{B})$ with respect to $\alpha_u^+$, and have 
    \begin{align}
    \label{equ:dev_alpha_plus}
        \frac{\partial P(\mathcal{B})}{\partial \alpha_u^+} 
         = \frac{\partial P(\mathcal{B})}{\partial U(\mathcal{B})}\cdot \frac{\partial U(\mathcal{B})}{\partial \alpha_u^+}
           + \frac{\partial P(\mathcal{B})}{\partial U(i_m)}\cdot \frac{\partial U(i_m)}{\partial \alpha_u^+}.
    \end{align}
    Based on Eq. \eqref{equ:dev_PB_UB} and Eq. \eqref{equ:dev_pb_um}, Eq. \eqref{equ:dev_alpha_plus} equals 
    \begin{align}
        \frac{\partial P(\mathcal{B})}{\partial \alpha_u^+}
        = P(\mathcal{B})P(i_m)\cdot \left\{\frac{\partial U(\mathcal{B})}{\partial \alpha_u^+} - \frac{\partial U(i_m)}{\partial \alpha_u^+}\right\}.
    \end{align}
    Since $U(i_m)$ is not a function of $\alpha_u^+$, we can focus on computing the derivatives of $U(\mathcal{B})$ with respect to $\alpha_u^+$. According to Eq. \eqref{equ:dev_ub_alpha_plus}, we have
    \begin{align}
        \frac{\partial P(\mathcal{B})}{\partial \alpha_u^+} 
        = \frac{1}{2}\cdot P(\mathcal{B}) P(i_m) u_1(c_\mathcal{B}) \cdot \ln \left[p(i_1^\mathcal{B}|i_m)\right]<0.
    \end{align}

    (2) Next, we consider the derivative of $P(\mathcal{B})$ with respect to $\alpha_u^-$. Similarly, we have 
    \begin{align}
        \frac{\partial P(\mathcal{B})}{\partial \alpha_u^-} 
        = P(\mathcal{B}) P(i_m)\cdot \left\{\frac{\partial U(\mathcal{B})}{\partial \alpha_u^-} - \frac{\partial U(i_m)}{\partial \alpha_u^-}\right\}. 
    \end{align}
    Since $U(\mathcal{B})$ is not a function of $\alpha_u^-$, we can focus on computing the derivatives of $U(i_m)$ with respect to $\alpha_u^-$. According to Eq. \eqref{equ:dev_um_alpha_minus}, we have
    \begin{align}
        \frac{\partial P(\mathcal{B})}{\partial \alpha_u^-} 
        = -\frac{1}{2}\cdot P(\mathcal{B}) P(i_m) u_1(c_{i_m}) \cdot \ln \left[p(\bar i_1^\mathcal{B}|i_m)\right]>0.
    \end{align}
\end{proof}


\textbf{Proof of Theorem 2.} The proof of Theorem 2 in the manuscript is as follows. 
\begin{proof}
    We consider the derivative of $P(\mathcal{B})$ with respect to $p(i_1^\mathcal{B}|i_m)$, and have 
    \begin{align}
    \label{equ:dev_pb_correlation}
        \frac{\partial P(\mathcal{B})}{\partial p(i_1^\mathcal{B}|i_m)} 
         = \frac{\partial P(\mathcal{B})}{\partial U(\mathcal{B})}\cdot \frac{\partial U(\mathcal{B})}{\partial p(i_1^\mathcal{B}|i_m)}
           + \frac{\partial P(\mathcal{B})}{\partial U(i_m)}\cdot \frac{\partial U(i_m)}{\partial p(i_1^\mathcal{B}|i_m)}.
    \end{align}
    Based on Eq. \eqref{equ:dev_PB_UB} and Eq. \eqref{equ:dev_pb_um}, Eq. \eqref{equ:dev_pb_correlation} equals 
    \begin{align}
    \label{equ:dev_corr_prob}
            \frac{\partial P(\mathcal{B})}{\partial p(i_1^\mathcal{B}|i_m)} 
            = P(\mathcal{B}) P(i_m) \cdot \left\{\frac{\partial U(\mathcal{B})}{\partial p(i_1^\mathcal{B}|i_m)} - \frac{\partial U(i_m)}{\partial p(i_1^\mathcal{B}|i_m)}\right\}.
    \end{align}
    To compute Eq. \eqref{equ:dev_corr_prob}, it is necessary to calculate the derivatives of $U(\mathcal{B})$ and $U(i_m)$ with respect to $p(i_1^\mathcal{B}|i_m)$, respectively.

    1) Firstly, for the derivatives of $U(\mathcal{B})$ with respect to $p(i_1^\mathcal{B}|i_m)$, we have
    \begin{align}
        \frac{\partial U(\mathcal{B})}{\partial p(i_1^\mathcal{B}|i_m)} = \frac{\partial \left[u_1(c_\mathcal{B})+u_2(\xi_{\mathcal{B}})\right]}{\partial p(i_1^\mathcal{B}|i_m)} = \frac{\partial u_1(c_\mathcal{B})}{\partial p(i_1^\mathcal{B}|i_m)}.
    \end{align}
    Recall that 
    \begin{align*}
        u_1(c_\mathcal{B}) = \left[p(i_1^\mathcal{B}|i_m)\right]^{(\alpha_u^+ + \alpha_{i_m}^+)/2}  \cdot v(c_m+c_1^\mathcal{B}-c_{\mathcal{B}}),
    \end{align*}
    then we have 
    \begin{align}
        \frac{\partial u_1(c_\mathcal{B})}{\partial p(i_1^\mathcal{B}|i_m)} 
        & = \frac{\alpha_u^+ + \alpha_{i_m}^+}{2}\cdot \left\{\left[p(i_1^\mathcal{B}|i_m)\right]^{(\alpha_u^+ + \alpha_{i_m}^+)/2-1}\right\} \\ \notag
        &\qquad \qquad \qquad \qquad \cdot v(c_m+c_1^\mathcal{B}-c_{\mathcal{B}})\\ \notag
        & = \frac{\alpha_u^+ + \alpha_{i_m}^+}{2}\cdot u_1(c_\mathcal{B}) \cdot  \left[p(i_1^\mathcal{B}|i_m)\right]^{-1}.
    \end{align}
    Therefore, 
    \begin{align}
    \label{equ:dev_correlation_probability_bundle}
            \frac{\partial U(\mathcal{B})}{\partial p(i_1^\mathcal{B}|i_m)} = \frac{\alpha_u^+ + \alpha_{i_m}^+}{2}\cdot u_1(c_\mathcal{B}) \cdot  \left[p(i_1^\mathcal{B}|i_m)\right]^{-1}.
    \end{align}

    2) Next, consider the derivatives of $U(i_m)$ with respect to $p(i_1^\mathcal{B}|i_m)$. similarly, we have 
    \begin{align}
        \frac{\partial U(i_m)}{\partial p(i_1^\mathcal{B}|i_m)} = \frac{\partial \left[u_1(c_{i_m})+u_2(\xi_{i_m})\right]}{\partial p(i_1^\mathcal{B}|i_m)} = \frac{\partial u_1(c_{i_m})}{\partial p(i_1^\mathcal{B}|i_m)}.
    \end{align}
    Recall that 
    \begin{align*}
        u_1(c_m) = \left[p(\bar i_1^\mathcal{B}|i_m)\right]^{(\alpha_u^- + \alpha_{i_m}^-)/2}\cdot v(c_{\mathcal{B}}-c_m),
    \end{align*}
    where $p(\bar i_1^\mathcal{B}|i_m)=1-p(i_1^\mathcal{B}|i_m)$, then we have 
    \begin{align}
        \frac{\partial u_1(c_{m})}{\partial p(i_1^\mathcal{B}|i_m)} 
        & = -\frac{\alpha_u^- + \alpha_{i_m}^-}{2}\cdot \left\{\left[1-p(i_1^\mathcal{B}|i_m)\right]^{(\alpha_u^- + \alpha_{i_m}^-)/2-1}\right\}   \\ \notag
        & \qquad \qquad \qquad \qquad \qquad \qquad \cdot v(c_{\mathcal{B}}-c_m)\\ \notag
        & = -\frac{\alpha_u^- + \alpha_{i_m}^-}{2}\cdot u_1(c_m) \cdot \left[1-p(i_1^\mathcal{B}|i_m)\right]^{-1}.
    \end{align}
    Therefore, 
    \begin{align}
    \label{equ:dev_correlation_probability_item}
        \frac{\partial U(i_m)}{\partial p(i_1^\mathcal{B}|i_m)} = -\frac{\alpha_u^- + \alpha_{i_m}^-}{2}\cdot u_1(c_m) \cdot \left[1-p(i_1^\mathcal{B}|i_m)\right]^{-1}.
    \end{align}

    Combining the results of Eq. \eqref{equ:dev_correlation_probability_bundle} and Eq. \eqref{equ:dev_correlation_probability_item}, we have
    \begin{small}
    \begin{align}
        \frac{\partial P(\mathcal{B})}{\partial p(i_1^\mathcal{B}|i_m)} 
            & = P(\mathcal{B}) P(i_m) \cdot
            \bigg\{\frac{\alpha_u^+ + \alpha_{i_m}^+}{2}\cdot u_1(c_\mathcal{B}) \cdot  \left[p(i_1^\mathcal{B}|i_m)\right]^{-1} \\ \notag
            & +\frac{\alpha_u^- + \alpha_{i_m}^-}{2}\cdot u_1(c_m) \cdot \left[1-p(i_1^\mathcal{B}|i_m)\right]^{-1}\bigg\}>0.
    \end{align}
    \end{small}
    That is, $P(\mathcal{B})$ is an increasing function of the correlation probability $p(i_1^\mathcal{B}|i_m)$.
\end{proof}

\textbf{Proof of Theorem 3.} The proof of Theorem 3 in the manuscript is as follows.
\begin{proof}
    We consider the derivative of $P(\mathcal{B})$ with respect to $c_1^\mathcal{B}$, and have
    \begin{align}
    \label{equ:dev_pb_c_additional}
        \frac{\partial P(\mathcal{B})}{\partial c_1^\mathcal{B}}
         = \frac{\partial P(\mathcal{B})}{\partial U(\mathcal{B})}\cdot \frac{\partial U(\mathcal{B})}{\partial c_1^\mathcal{B}}
           + \frac{\partial P(\mathcal{B})}{\partial U(i_m)}\cdot \frac{\partial U(i_m)}{\partial c_1^\mathcal{B}}.
    \end{align}
    Based on Eq. \eqref{equ:dev_PB_UB} and Eq. \eqref{equ:dev_pb_um}, Eq. \eqref{equ:dev_pb_c_additional} equals
    \begin{align}
    \label{equ:dev_pb_c_addtion}
            \frac{\partial P(\mathcal{B})}{\partial c_1^\mathcal{B}}
            &= P(\mathcal{B}) P(i_m) \cdot \left\{\frac{\partial U(\mathcal{B})}{\partial c_1^\mathcal{B}} - \frac{\partial U(i_m)}{\partial c_1^\mathcal{B}}\right\} \\ \notag 
            &= P(\mathcal{B}) P(i_m) \cdot \left\{\frac{\partial u_1(c_\mathcal{B})}{\partial c_1^\mathcal{B}} - \frac{\partial u_1(c_m)}{\partial c_1^\mathcal{B}}\right\}
    \end{align}
    
    1) We first compute the value of ${\partial P(\mathcal{B})}\big /{\partial c_1^\mathcal{B}}$. To compute Eq. \eqref{equ:dev_pb_c_addtion}, it is necessary to calculate the derivatives of $u_1(c_\mathcal{B})$ and $u_1(c_{i_m})$ with respect to $c_1^\mathcal{B}$, respectively. \\
    1.1) We begin with the derivatives of $u_1(c_\mathcal{B})$ with respect to $c_1^\mathcal{B}$. Recall that 
    \begin{align*}
        u_1(c_\mathcal{B}) = \left[p(i_1^\mathcal{B}|i_m)\right]^{(\alpha_u^+ + \alpha_{i_m}^+)/2}  \cdot v(c_m+c_1^\mathcal{B}-c_{\mathcal{B}}),
    \end{align*}
    where $c_{\mathcal{B}}=r\cdot(c_m+c_1^\mathcal{B})$ and $v(x)=x^{\beta^+}$. Then, we have  
    \begin{align*}
        u_1(c_\mathcal{B}) = \left[p(i_1^\mathcal{B}|i_m)\right]^{(\alpha_u^+ + \alpha_{i_m}^+)/2}  \cdot \left[(1-r)\cdot(c_m+c_1^\mathcal{B})\right]^{\beta^+},
    \end{align*}
    and
    \begin{align}
    \label{equ:dev_c_bundle}
        \frac{\partial u_1(c_\mathcal{B})}{\partial c_1^\mathcal{B}} 
        &= \left[p(i_1^\mathcal{B}|i_m)\right]^{(\alpha_u^+ + \alpha_{i_m}^+)/2} \cdot \beta^+ \\ \notag
        &\qquad \qquad \qquad \cdot \left[(1-r)\cdot(c_m+c_1^\mathcal{B})\right]^{\beta^+-1}\cdot (1-r).
    \end{align}
    1.2) Consider the derivatives of $u_1(c_{i_m})$ with respect to $c_1^\mathcal{B}$. Recall that 
    \begin{align*}
        u_1(c_m) &= \left[p(\bar i_1^\mathcal{B}|i_m)\right]^{(\alpha_u^- + \alpha_{i_m}^-)/2}\cdot v(c_{\mathcal{B}}-c_m) \\ \notag
        &= \left[p(\bar i_1^\mathcal{B}|i_m)\right]^{(\alpha_u^- + \alpha_{i_m}^-)/2}\cdot \left[r\cdot(c_m+c_1^\mathcal{B})-c_m\right]^{\beta^+}
    \end{align*}
    Then,  
    \begin{align}
    \label{equ:dev_c_item}
        \frac{\partial u_1(c_{i_m})}{\partial c_1^\mathcal{B}} 
        &= \left[p(\bar i_1^\mathcal{B}|i_m)\right]^{(\alpha_u^- + \alpha_{i_m}^-)/2}\cdot \beta^+ \\ \notag
        &\qquad \qquad \cdot \left[r\cdot(c_m+c_1^\mathcal{B})-c_m\right]^{\beta^+ - 1}\cdot r.
    \end{align}
    Combining the results of Eq. \eqref{equ:dev_c_bundle} and Eq. \eqref{equ:dev_c_item}, we have 
    \begin{align}
    \label{equ:dev_c_addition_second}
        &\frac{\partial P(\mathcal{B})}{\partial c_1^\mathcal{B}} 
             = \beta^+ \cdot P(\mathcal{B}) P(i_m) \cdot\bigg\{ \\ \notag
            & \left[p(i_1^\mathcal{B}|i_m)\right]^{(\alpha_u^+ + \alpha_{i_m}^+)/2} \cdot \left[(1-r)\cdot(c_m+c_1^\mathcal{B})\right]^{\beta^+-1}\cdot (1-r)  \\ \notag
            & -\left[p(\bar i_1^\mathcal{B}|i_m)\right]^{(\alpha_u^- + \alpha_{i_m}^-)/2}\cdot \left[r\cdot(c_m+c_1^\mathcal{B})-c_m\right]^{\beta^+ - 1}\cdot r\bigg\}.
    \end{align}

    (2) Next, we analyze the equivalent conditions for ${\partial P(\mathcal{B})}/{\partial c_1^\mathcal{B}}>0$. To determine these conditions, we examine the following inequality that needs to be satisfied.
    \begin{align}
    \label{equ:dev_c_addition_third}
        &\left[p(i_1^\mathcal{B}|i_m)\right]^{(\alpha_u^+ + \alpha_{i_m}^+)/2} \cdot \left[(1-r)\cdot(c_m+c_1^\mathcal{B})\right]^{\beta^+-1} \cdot (1-r) \\ \notag
        &\; > \; \left[p(\bar i_1^\mathcal{B}|i_m)\right]^{(\alpha_u^- + \alpha_{i_m}^-)/2}\cdot \left[r\cdot(c_m+c_1^\mathcal{B})-c_m\right]^{\beta^+ - 1}\cdot r  \\ \notag
        &\Leftrightarrow \qquad \left[\frac{c_m+c_1^\mathcal{B}}{r\cdot(c_m+c_1^\mathcal{B})-c_m}\right]^{\beta^+-1} \\ \notag
        &\qquad \qquad \qquad \qquad  >\; \frac{\left[p(\bar i_1^\mathcal{B}|i_m)\right]^{(\alpha_u^- + \alpha_{i_m}^-)/2}}{\left[p(i_1^\mathcal{B}|i_m)\right]^{(\alpha_u^+ + \alpha_{i_m}^+)/2}}\cdot \frac{r}{(1-r)^{\beta^+}}.  \\ \notag
        &\Leftrightarrow \qquad \left[\frac{r\cdot(c_m+c_1^\mathcal{B})-c_m}{c_m+c_1^\mathcal{B}}\right]^{1-\beta^+} \\ \notag
        &\qquad \qquad \qquad \qquad  >\; \frac{\left[p(\bar i_1^\mathcal{B}|i_m)\right]^{(\alpha_u^- + \alpha_{i_m}^-)/2}}{\left[p(i_1^\mathcal{B}|i_m)\right]^{(\alpha_u^+ + \alpha_{i_m}^+)/2}}\cdot \frac{r}{(1-r)^{\beta^+}} \notag
    \end{align}
    As $1-\beta^+>0$, Eq. \eqref{equ:dev_c_addition_third} equals
    \begin{align}
    \label{equ:dev_c_addition_forth}
        &\frac{r\cdot(c_m+c_1^\mathcal{B})-c_m}{c_m+c_1^\mathcal{B}} \\ \notag
        &\qquad >\;\sqrt[1-\beta^+]{\frac{\left[p(\bar i_1^\mathcal{B}|i_m)\right]^{(\alpha_u^- + \alpha_{i_m}^-)/2}}{\left[p(i_1^\mathcal{B}|i_m)\right]^{(\alpha_u^+ + \alpha_{i_m}^+)/2}}\cdot \frac{r}{(1-r)^{\beta^+}}} \\ \notag
        \Leftrightarrow \quad& \frac{r\cdot(c_m+c_1^\mathcal{B})-c_m}{c_m+c_1^\mathcal{B}} \\ \notag
        &\qquad >\;\sqrt[1-\beta^+]{\frac{\left[p(\bar i_1^\mathcal{B}|i_m)\right]^{(\alpha_u^- + \alpha_{i_m}^-)/2}}{\left[p(i_1^\mathcal{B}|i_m)\right]^{(\alpha_u^+ + \alpha_{i_m}^+)/2}}}\cdot 
        \sqrt[1-\beta^+]{\frac{r}{(1-r)^{\beta^+}} }  \notag
    \end{align}
    Let 
    \begin{align}
        \mathcal{A} \coloneqq \sqrt[1-\beta^+]{\frac{\left[p(\bar i_1^\mathcal{B}|i_m)\right]^{(\alpha_u^- + \alpha_{i_m}^-)/2}}{\left[p(i_1^\mathcal{B}|i_m)\right]^{(\alpha_u^+ + \alpha_{i_m}^+)/2}}}>0,
    \end{align}
    then Eq. \eqref{equ:dev_c_addition_forth} equals
    \begin{align}
    \label{equ:dev_c_addition_fifth}
        &\frac{r\cdot(c_m+c_1^\mathcal{B})-c_m}{c_m+c_1^\mathcal{B}} > \mathcal{A}\cdot \sqrt[1-\beta^+]{\frac{r}{(1-r)^{\beta^+}} } \\  \notag
        \Leftrightarrow \quad &r\cdot(c_m+c_1^\mathcal{B})-c_m > \mathcal{A}\cdot \sqrt[1-\beta^+]{\frac{r}{(1-r)^{\beta^+}} }\cdot (c_m+c_1^\mathcal{B})  \\ \notag
        \Leftrightarrow  \quad &\left[r-\mathcal{A}\cdot \sqrt[1-\beta^+]{\frac{r}{(1-r)^{\beta^+}} }\right]\cdot c_1^\mathcal{B} \\ \notag
        &\qquad \qquad \qquad >\; \left[(1-r)+\mathcal{A}\cdot \sqrt[1-\beta^+]{\frac{r}{(1-r)^{\beta^+}} }\right]\cdot c_m \\ \notag
        \Leftrightarrow \quad & r\cdot \left[1-\mathcal{A}\cdot \sqrt[1-\beta^+]{\frac{r}{(1-r)^{\beta^+}} }\cdot r^{-1}\right]\cdot c_1^\mathcal{B} \\ \notag
        &\qquad \qquad \qquad>\; \left[(1-r)+\mathcal{A}\cdot \sqrt[1-\beta^+]{\frac{r}{(1-r)^{\beta^+}} }\right]\cdot c_m \notag
    \end{align}
    As
    \begin{align}
    \label{equ:r_substitute}
        &\sqrt[1-\beta^+]{\frac{r}{(1-r)^{\beta^+}} }\cdot r^{-1} = r^{\frac{1}{1-\beta^+}}\cdot\left(\frac{1}{1-r}\right)^{\frac{\beta^+}{1-\beta^+}}\cdot r^{-1} \\ \notag
        &= r^{\frac{\beta^+}{1-\beta^+}}\cdot \left(\frac{1}{1-r}\right)^{\frac{\beta^+}{1-\beta^+}} = \left(\frac{r}{1-r}\right)^{\frac{\beta^+}{1-\beta^+}},  \notag
    \end{align}
    then Eq. \eqref{equ:dev_c_addition_fifth} equals 
    \begin{align}
    \label{equ:dev_c_addition_sixth}
        &r\cdot \left[1-\mathcal{A}\cdot \left(\frac{r}{1-r}\right)^{\frac{\beta^+}{1-\beta^+}}\right]\cdot c_1^\mathcal{B}   \\ \notag
        &\qquad \qquad \qquad>\; \left[(1-r)+\mathcal{A}\cdot \sqrt[1-\beta^+]{\frac{r}{(1-r)^{\beta^+}} }\right]\cdot c_m.
    \end{align}
    That is, the equivalent conditions for ${\partial P(\mathcal{B})}\big /{\partial c_1^\mathcal{B}}>0$ is that Eq. \eqref{equ:dev_c_addition_sixth} holds true.
    
    3) Finally, we analyze the conditions that need to be satisfied for $c_1^\mathcal{B}$ in order for Eq. \eqref{equ:dev_c_addition_sixth} to hold. Define 
    \begin{align}
        \kappa \coloneqq \frac{\left[(1-r)+\mathcal{A}\cdot \sqrt[1-\beta^+]{\frac{r}{(1-r)^{\beta^+}} }\right]}
        {r\cdot \left[1-\mathcal{A}\cdot \left(\frac{r}{1-r}\right)^{\frac{\beta^+}{1-\beta^+}}\right]}.
    \end{align}
    Since the inequality
    \begin{align}
        \left[(1-r)+\mathcal{A}\cdot \sqrt[1-\beta^+]{\frac{r}{(1-r)^{\beta^+}} }\right]>0
    \end{align}
    always holds, we can conclude that $\kappa>0$ when 
    \begin{align}
        r\cdot \left[1-\mathcal{A}\cdot \left(\frac{r}{1-r}\right)^{\frac{\beta^+}{1-\beta^+}}\right]>0.
    \end{align}
    Notice that  
    \begin{align}
    \label{equ:dev_c_addition_seventh}
        &1-\mathcal{A}\cdot \left(\frac{r}{1-r}\right)^{\frac{\beta^+}{1-\beta^+}}>0 \\ \notag \Leftrightarrow \quad 
        &\left(\frac{r}{1-r}\right)^{\frac{\beta^+}{1-\beta^+}} < \frac{1}{\mathcal{A}} \\ \notag
        \Leftrightarrow \quad 
        &\frac{r}{1-r} < \mathcal{A}^{-\frac{1-\beta^+}{\beta^+}} \\ \notag
        \Leftrightarrow \quad 
        &\left(1+\mathcal{A}^{-\frac{1-\beta^+}{\beta^+}}\right)\cdot r < \mathcal{A}^{-\frac{1-\beta^+}{\beta^+}} \\ \notag
        \Leftrightarrow \quad 
        &r<\frac{\mathcal{A}^{-\frac{1-\beta^+}{\beta^+}}}{1+\mathcal{A}^{-\frac{1-\beta^+}{\beta^+}}} = \frac{1}{1+\mathcal{A}^{\frac{1-\beta^+}{\beta^+}}}.
    \end{align}
    That is, we have $\kappa>0$ when $r<1\big /\left(1+\mathcal{A}^{\frac{1-\beta^+}{\beta^+}}\right)$, and $\kappa<0$ otherwise.

    Combining the results of Eq. \eqref{equ:dev_c_addition_sixth} and Eq. \eqref{equ:dev_c_addition_seventh}, the following conclusions can be drawn. 
    \begin{itemize}
        \item If $r<1\big /\left(1+\mathcal{A}^{\frac{1-\beta^+}{\beta^+}}\right)$, then we have $\kappa>0$, and 
        \begin{align}
            &\frac{\partial P(\mathcal{B})}{\partial c_1^\mathcal{B}}>0 \quad \text{when} \quad c_1^\mathcal{B} > \kappa\cdot c_m, \\ \notag
            \text{and} \quad &\frac{\partial P(\mathcal{B})}{\partial c_1^\mathcal{B}}<0 \quad \text{when} \quad c_1^\mathcal{B} < \kappa\cdot c_m.
        \end{align}
        \item If $r>1\big /\left(1+\mathcal{A}^{\frac{1-\beta^+}{\beta^+}}\right)$, then we have $\kappa<0$, and 
        \begin{align}
            &\frac{\partial P(\mathcal{B})}{\partial c_1^\mathcal{B}}>0 \quad \text{when} \quad c_1^\mathcal{B} < \kappa\cdot c_m, \\ \notag
            \text{and} \quad &\frac{\partial P(\mathcal{B})}{\partial c_1^\mathcal{B}}<0 \quad \text{when} \quad  c_1^\mathcal{B} > \kappa\cdot c_m.
        \end{align}
        As $\kappa<0$ and $\kappa\cdot c_m<0$, then we have
        \begin{align}
            \frac{\partial P(\mathcal{B})}{\partial c_1^\mathcal{B}}<0 \quad \text{when} \quad  c_1^\mathcal{B}\in (0,+\infty)
        \end{align} 
    \end{itemize}
    
\end{proof}

\subsection{Experiments}

\subsubsection{Experimental Setup}


Among the baselines considered in our work, binary classification-based methods and learning-to-rank methods require numerical feature vectors as input. To compare with these methods, we generate features for users, individual games, and bundles based on the following step. 
\begin{itemize}
    \item For an individual game $i_m$, both direct and indirect features can be considered. The direct features include attributes such as price, genre, average playing time, the number of reviews, and the number of recommendations for the game. The indirect features capture the ratios of recommendations to reviews and reviews to purchases for the game. Combining these features, an individual game $i_m$ can be represented as a 27-dimensional vector $\boldsymbol{v}_{i_m}$.
    \item For a bundle ${\mathcal{B}}$, its feature representation can be obtained as the average feature vector of the games contained in the bundle. In other words, $\boldsymbol{v}_{\mathcal{B}}$ is calculated as the average of the feature vectors $\boldsymbol{v}_{i_j}$ for each game $i_j$ in the bundle.
    \item For a user $u$, the feature representation can be obtained by averaging the features of the games that the user has purchased. That is, $\boldsymbol{v}_{u}$ is calculated as the average of the feature vectors $\boldsymbol{v}_{i_j}$ for each game $i_j$ in the set $N_u$, where $|N_u|$ denotes the total number of games in the user's purchase history.
\end{itemize}
By using these feature representations, individual games, bundles, and users can be represented as numerical vectors, allowing binary classification-based methods and learning-to-rank methods to be applied and compared based on these aggregated features.

In this paper, we compare the proposed \emph{Probe} with the following baselines. Detailed implementation of these methods are as follows. 
\begin{itemize}
    \item Frequency-based method. During the training stage, we calculate the number of bundle purchases $M(\mathcal{B})$ and the number of single item purchases $M(i_m)$ for each user, and compute the ratio $p = M(\mathcal{B}) / M(i_m)$. During the testing stage, we assume that the will select a bundle with the probability of $p$, and will choose a single game with the probability of $1-p$.
    \item Binary Classification-Based Method. In this approach, we employ binary classifiers to determine the probability of each user's purchase record being labeled as selecting the bundle. To facilitate the classification process, we utilize a feature vector $\boldsymbol{v} = [\boldsymbol{v}_u, \boldsymbol{v}_{i_m}, \boldsymbol{v}_b]$ as input to the binary classifiers. If the input vector is labeled as the positive, we consider that the user will select the bundle;  otherwise, we consider the user will select the single item. To compare the performance of various classifiers, we evaluate the Naive Bayes classifier, Support Vector Machines (SVM), Gradient Boosting Decision Trees (GBDT), AdaBoost, and other classifiers. Among these options, we find that AdaBoost achieves the best performance based on the evaluation metrics employed. Therefore, we present the results obtained using AdaBoost as an illustrative example of the binary classification-based method.
    \item Learning-to-rank methods. For RankSVM, during the training stage, it utilizes the vector $\boldsymbol{v} = [\boldsymbol{v}_b-\boldsymbol{v}_{i_m}, \boldsymbol{v}_u]$ as input to train a model that captures the decision boundary based on users' preferences. In the testing stage, RankSVM takes the input vector $\boldsymbol{v}$ and generates a real-valued output. If the output value falls on the right side of the decision boundary, it is assumed that users will choose the bundle. Conversely, if the output value falls on the left side of the decision boundary, it is assumed that users will choose a single game. For RankNet, it takes two inputs: $\boldsymbol{v}_1 = [\boldsymbol{v}_b, \boldsymbol{v}_u]$ and $\boldsymbol{v}_2 = [\boldsymbol{v}_{i_m}, \boldsymbol{v}_u]$. The model's output is the probability that users prefer $\boldsymbol{v}_1$ (the bundle) over $\boldsymbol{v}_2$ (the individual game). If the probability is greater than 0.5, it is assumed that users will choose the bundle. Conversely, if the probability is less than or equal to 0.5, it is assumed that users will choose a single game.
    \item Bundle recommendation methods. During the training stage, we utilize these bundle recommendation methods, such as BPR methods, BGCN, and CrossCBR, to generate the vector representations for users, individual games, and bundles. During the testing stage, we calculates the similarity between the representations of users and individual games, as well as between users and bundles. The similarity is then translated into the probability of users choosing single games or choosing bundles using the softmax function. 
\end{itemize}

\subsubsection{Experimental Results}

\begin{figure}[tbp]
\centering
\begin{minipage}[b]{0.45\linewidth}
  \centering
  \centerline{\includegraphics[width=4.2cm]{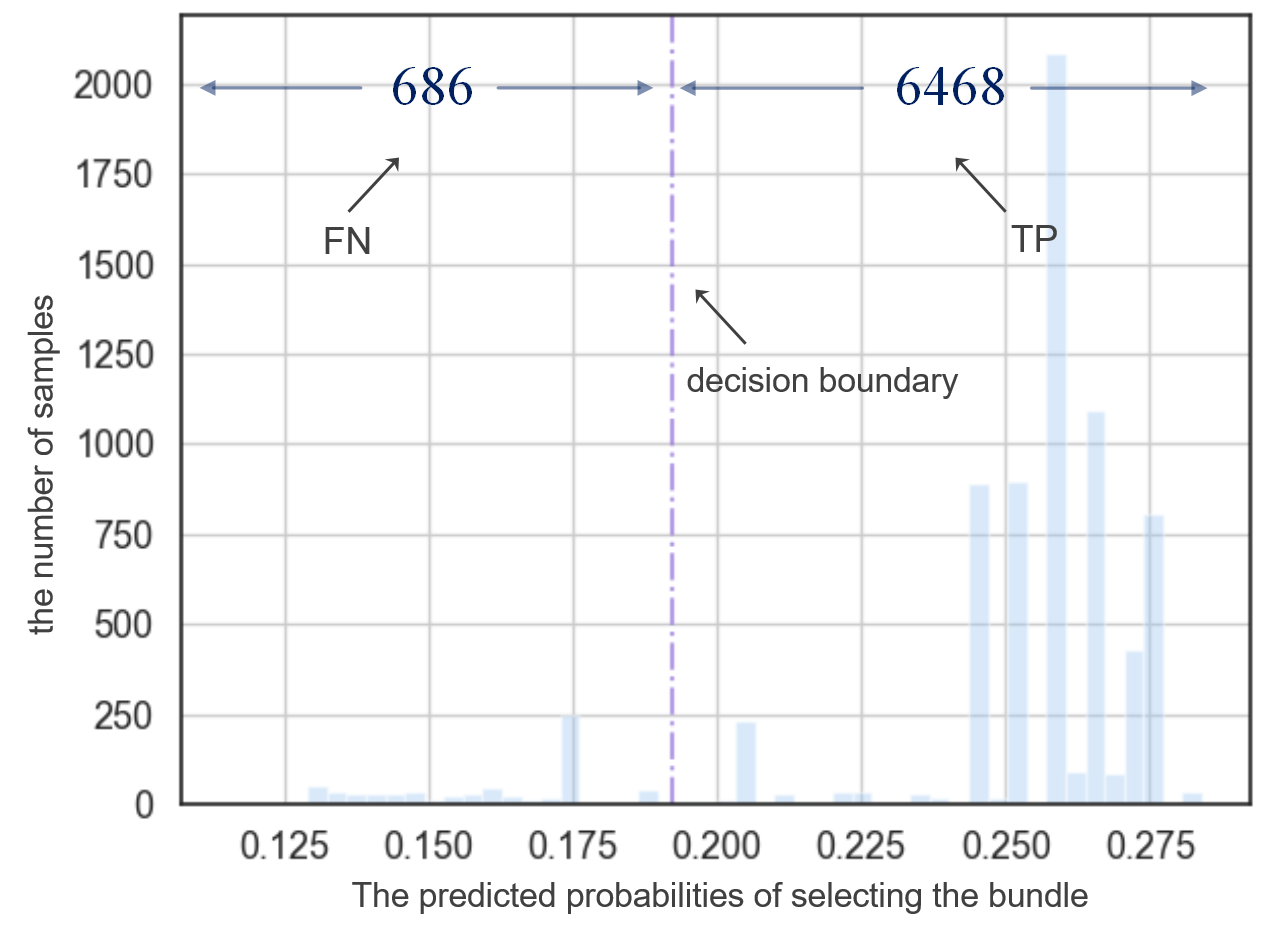}}
  \centerline{(a)}\medskip
\end{minipage}
\hfill
\begin{minipage}[b]{0.45\linewidth}
  \centering
  \centerline{\includegraphics[width=4.2cm]{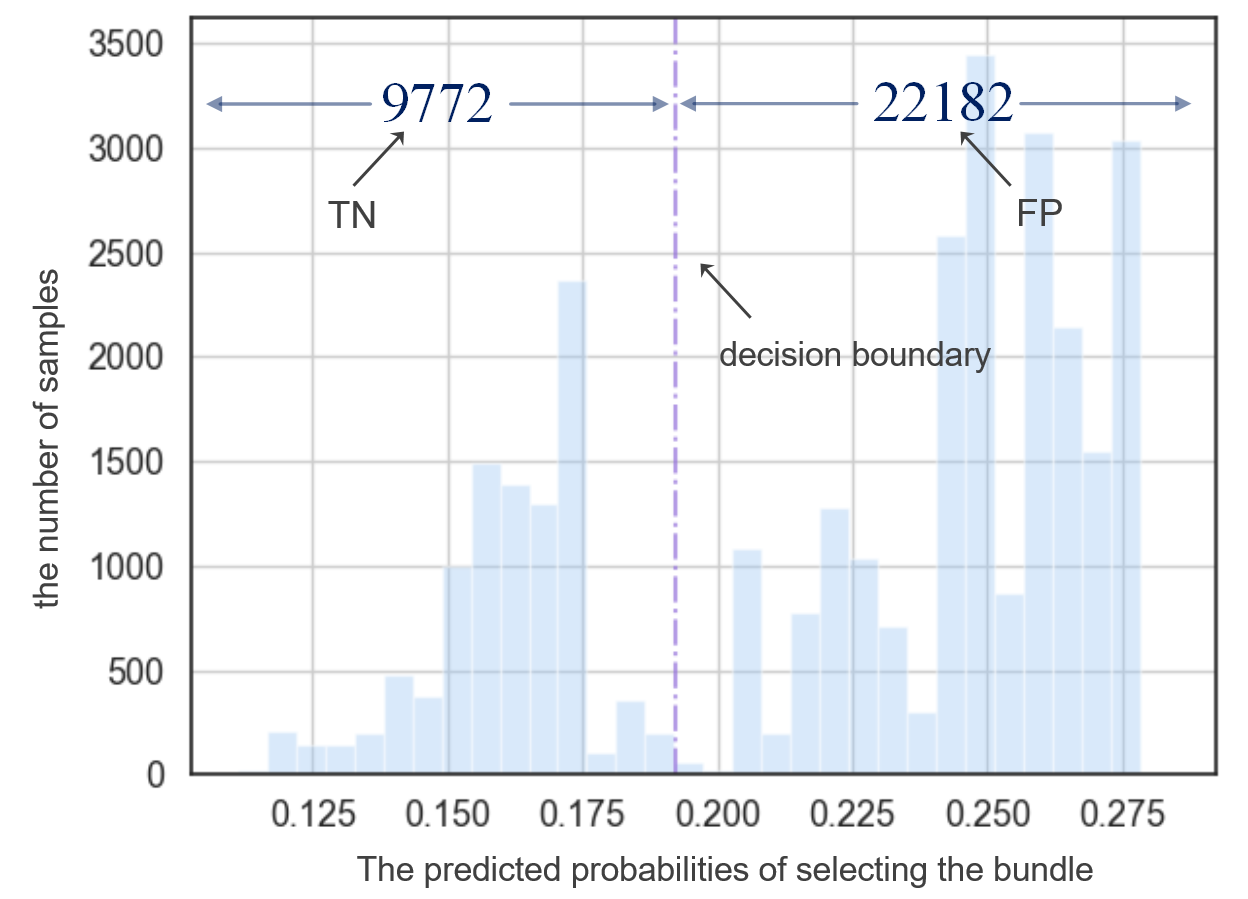}}
  \centerline{(b)}\medskip
\end{minipage}
\hfill
\begin{minipage}[b]{0.45\linewidth}
  \centering
  \centerline{\includegraphics[width=4.2cm]{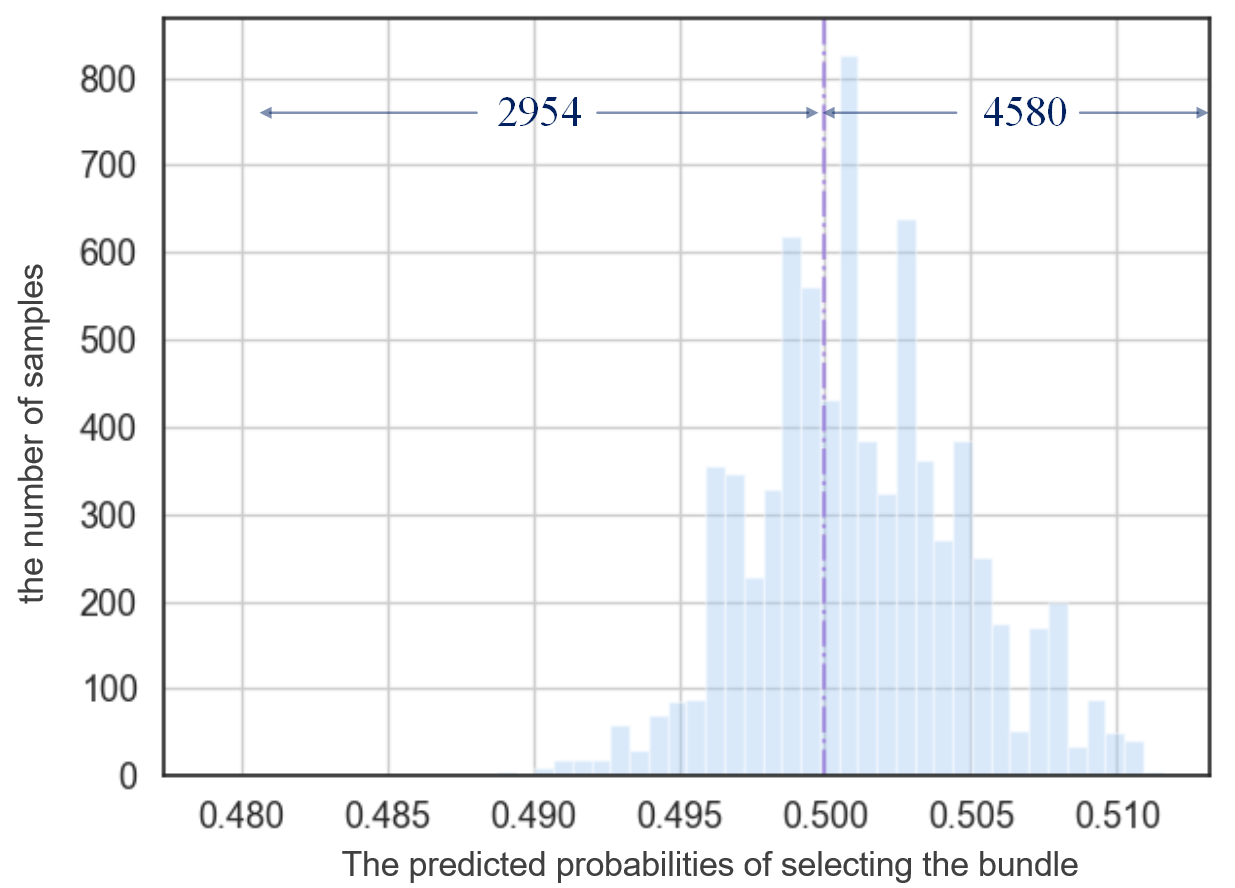}}
  \centerline{(c) }\medskip
\end{minipage}
\hfill
\begin{minipage}[b]{0.45\linewidth}
  \centering
  \centerline{\includegraphics[width=4.2cm]{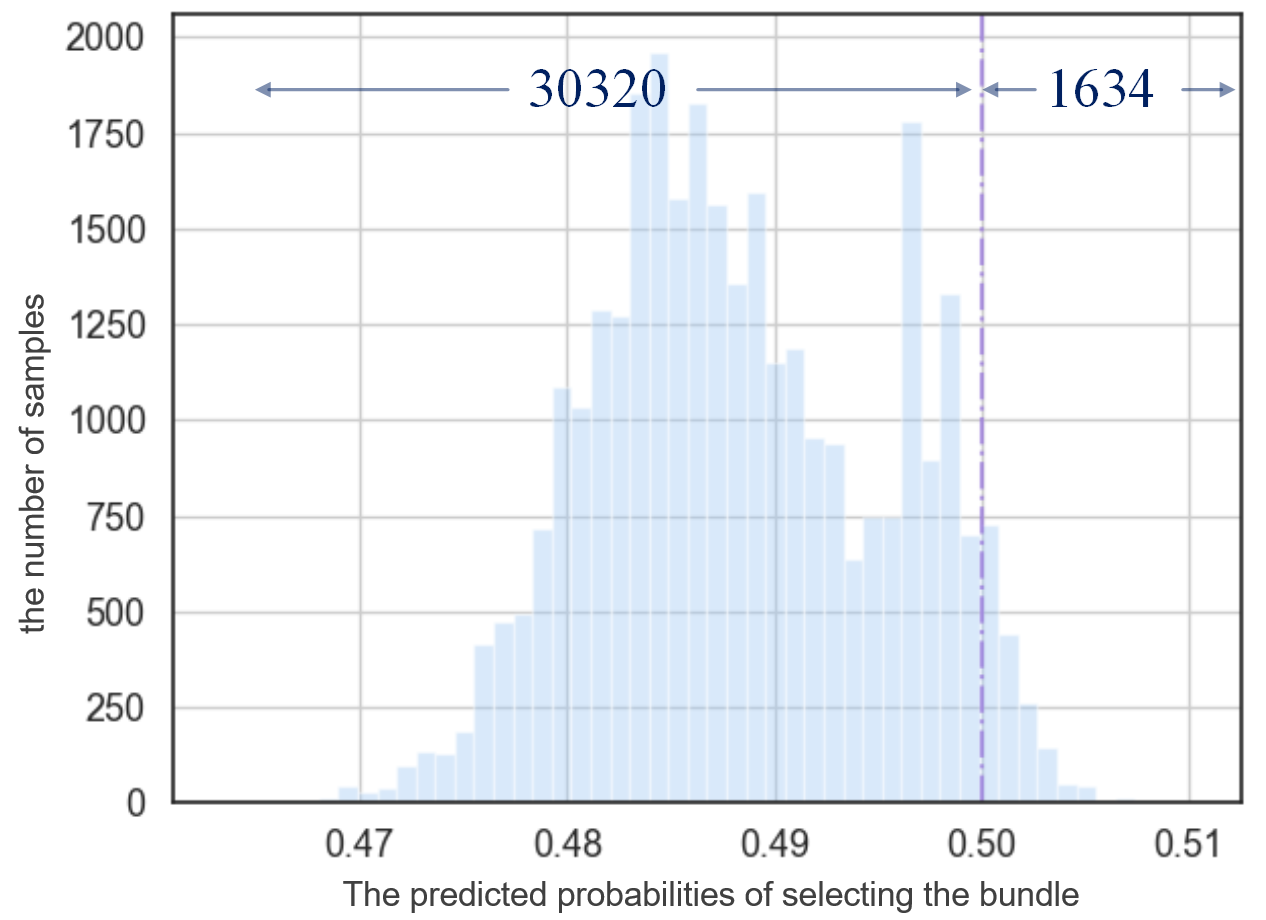}}
  \centerline{(d)}\medskip
\end{minipage}
\hfill
\begin{minipage}[b]{0.45\linewidth}
  \centering
  \centerline{\includegraphics[width=4.2cm]{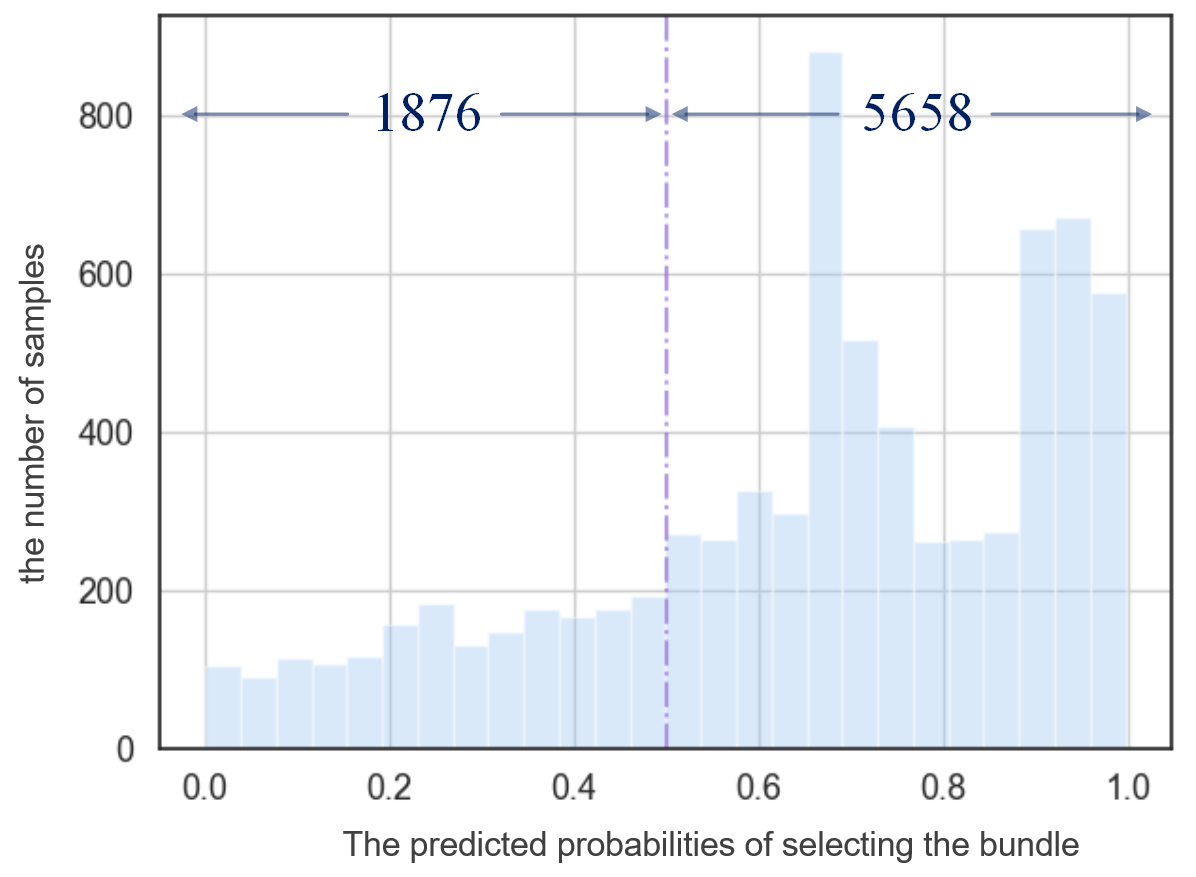}}
  \centerline{(e)}\medskip
\end{minipage}
\hfill
\begin{minipage}[b]{0.45\linewidth}
  \centering
  \centerline{\includegraphics[width=4.2cm]{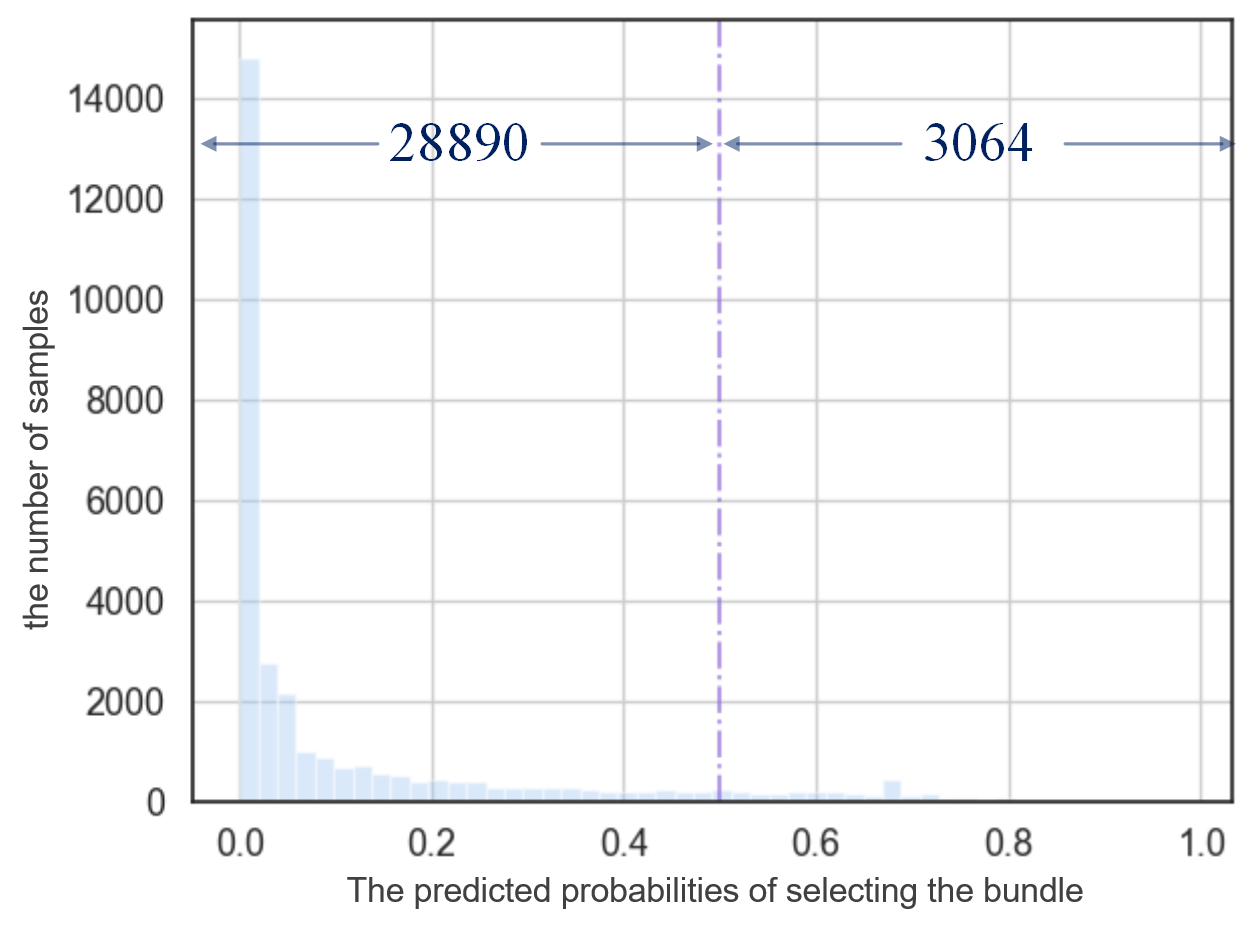}}
  \centerline{(f)}\medskip
\end{minipage}
\hfill
\begin{minipage}[b]{0.45\linewidth}
  \centering
  \centerline{\includegraphics[width=4.2cm]{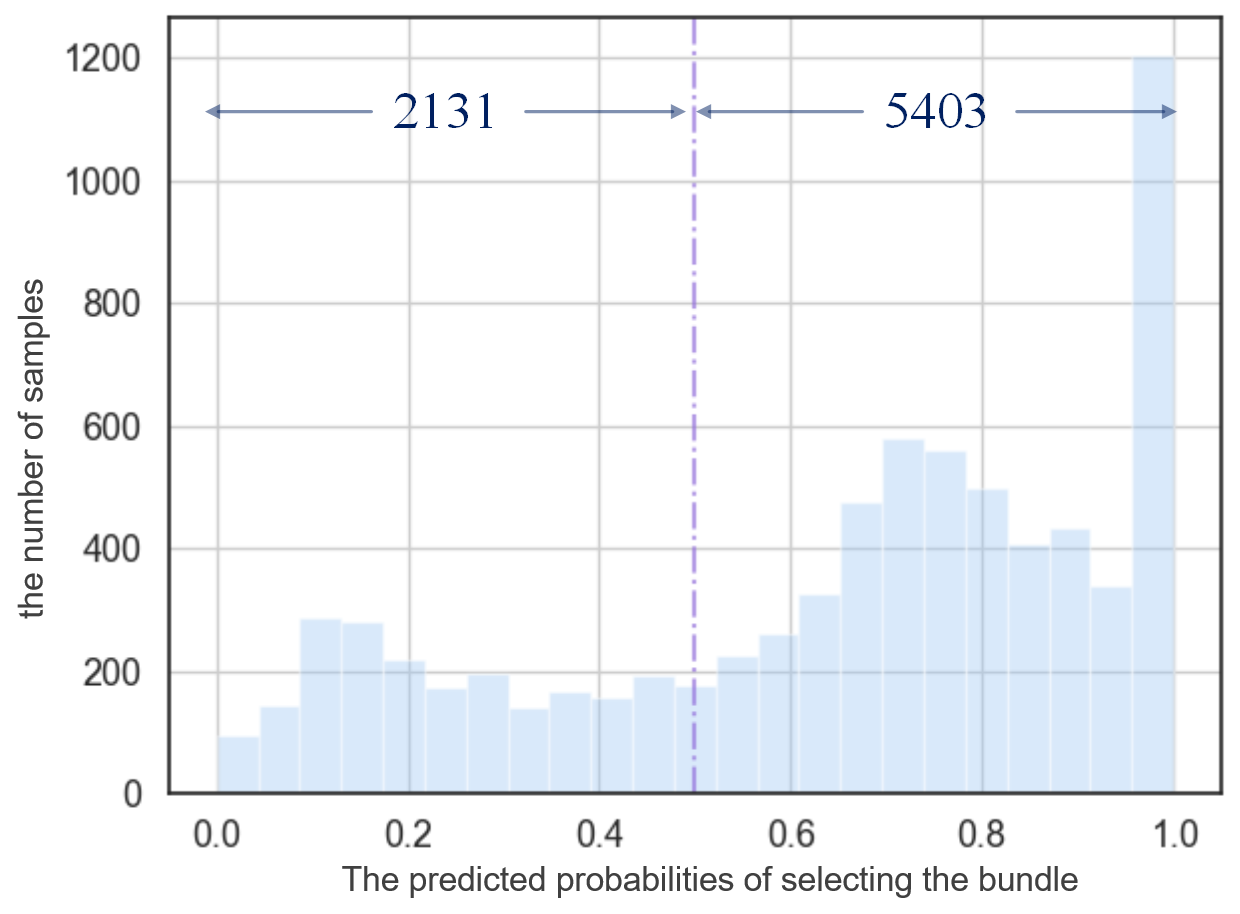}}
  \centerline{(g)}\medskip
\end{minipage}
\hfill
\begin{minipage}[b]{0.45\linewidth}
  \centering
  \centerline{\includegraphics[width=4.2cm]{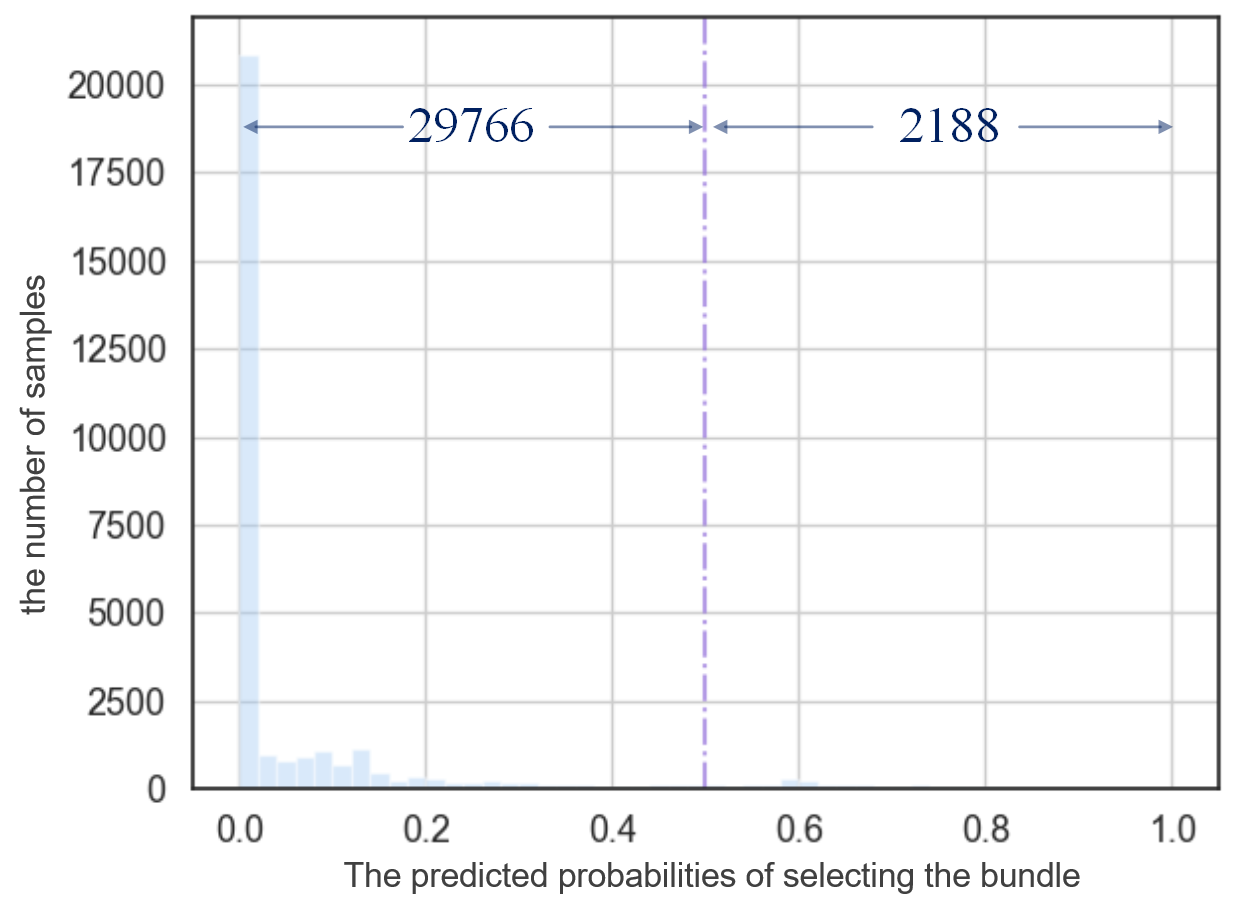}}
  \centerline{(h)}\medskip
\end{minipage}
\caption{The comparison of the predicted probabilities. (a) RankSVM, where users actually choose the bundle. (b) RankSVM, where users actually choose a single game. (c) AdaBoost, where users actually choose the bundle. (d) AdaBoost, where users actually choose a single game. (e) RankNet, where users actually choose the bundle. (f) RankNet, where users actually choose a single game. (g) \emph{Probe}, where users actually choose the bundle. (h) \emph{Probe}, where users actually choose a single game.}
\label{fig:accu_prob}
\end{figure}

To explain the simulation results in Table III in the manuscript, we plot the histogram of the predicted $P(i_m)$ and $P(B)$ for \emph{Probe}, RankNet, RankSVM, and AdaBoost. Specifically, Fig.\ref{fig:accu_prob} (a) and (b) display the predicted probabilities $P_{select}(\mathcal{B})$ obtained by the RankSVM method. Fig.\ref{fig:accu_prob} (a) corresponds to cases where users actually select the bundle, while Fig.\ref{fig:accu_prob} (b) corresponds to cases where users actually select a single game. In Fig.\ref{fig:accu_prob} (a), the purple dashed line represents the decision boundary, and the records on the right side of the decision boundary are correctly predicted. Similarly, in Fig.\ref{fig:accu_prob} (b), the records on the left side of the decision boundary are correctly predicted. 

Based on the observations from Fig.~\ref{fig:accu_prob} (b), it can be noted that the RankSVM method incorrectly predicts a significant number of records where users actually choose a single game as choosing the bundle. This indicates that the RankSVM method has low precision. Similarly, according to Fig.~\ref{fig:accu_prob} (c), the AdaBoost method incorrectly predicts a large portion of records where users actually choose a bundle as choosing a single game. As a result, the AdaBoost method has low recall. Additionally, the predicted probabilities of the RankSVM and AdaBoost methods are concentrated near the decision boundary, leading to their overall poor performance in terms of the $F1$ metric.

On the other hand, Fig.\ref{fig:accu_prob} (e) indicates that most of the predicted probabilities generated by the RankNet method are greater than 0.5 when users actually select the bundle. Conversely, according to Fig.\ref{fig:accu_prob} (f), the predicted probabilities of the RankNet method are close to 0 when users actually select a single game. These observations suggest that the predicted probabilities of the RankNet method are relatively distant from the decision boundary, resulting in RankNet achieving the highest $F1$ metric among the baseline methods.

Furthermore, considering the proposed \emph{Probe}, Fig.\ref{fig:accu_prob} (g) demonstrates a reduction of 255 correctly predicted records where users choose the bundle compared to RankNet (5658 - 5403). Moreover, Fig.\ref{fig:accu_prob} (h) shows an increase of 876 correctly predicted records where users choose a single game compared to RankNet (29766 - 28890). Additionally, the predicted probabilities generated by \emph{Probe} for both types of records are farther away from the decision boundary than those of RankNet. Consequently, the proposed \emph{Probe} achieves a $2\%$ improvement in the $F1$ metric compared to RankNet.

The analysis reveals that the predicted probabilities of RankSVM and AdaBoost are concentrated near the decision boundary, which explains their relatively poor overall performance in terms of the $F1$ metric. In contrast, the predicted probabilities of RankNet are well-separated from the decision boundary, resulting in its higher $F1$ metric compared to other baseline methods. However, the predicted probabilities of the proposed \emph{Probe} method are even farther away from the decision boundary than those of RankNet. As a result, the proposed \emph{Probe} method still achieves a $2\%$ improvement in the $F1$ metric compared to RankNet.The results presented in Fig.\ref{fig:accu_prob} illustrate the predicted probabilities of users selecting the bundle for the different methods on the testing dataset.

\end{document}